\documentclass{elsart}
\usepackage{graphicx}
\usepackage{epsfig}
\usepackage{amssymb}
\usepackage{bbm}
\usepackage{subfigure}

\newcommand{\Z}{\mathbb{Z}}

\newcommand{\R}{\mathbb{R}}
\newcommand{\C}{\mathbb{C}}
\newcommand{\tensor}{\otimes}
\newcommand{\tens}{\otimes}
\newcommand{\phir}{\phi_r}
\newcommand{\phil}{\phi_l}
\newcommand{\te}{\theta}
\newcommand{\sd}{\rtimes}
\newcommand{\Rcal}{\mathcal{T}_r}
\newcommand{\Tcal}{\mathcal{T}}
\newcommand{\Acal}{\mathcal{A}}
\newcommand{\Ucal}{\mathcal{U}}
\newcommand{\Ccal}{\mathcal{C}}
\newcommand{\Lcal}{\mathcal{T}_l}
\newcommand{\Ical}{\mathcal{I}}

\newcommand{\Tdo}{\tilde{T}}

\newcommand{\braid}{\mathcal{R}}
\newcommand{\eps}{\varepsilon}

\begin{document}

\begin{frontmatter}

\title{The breaking of quantum double symmetries by defect condensation}

\author{F.A. Bais}, \ead{bais@science.uva.nl}
\author{C.J.M. Mathy\thanksref{now}}
\thanks[now]{Present address: Physics Department, Princeton University,
Jadwin Hall, Princeton, NJ 08544, United States}
\ead{cmathy@princeton.edu}

\address{Institute for Theoretical Physics,
University of Amsterdam, Valckenierstraat 65, 1018 XE Amsterdam, The
Netherlands}

\begin{abstract}
\noindent In this paper, we study the phenomenon of Hopf or more
specifically  quantum double symmetry breaking. We devise a
criterion for this type of symmetry breaking which is more general
than the one originally proposed in \cite{Bais:2002pb,Bais:2002ny},
and therefore extends the number of possible breaking patterns that
can be described consistently. We start by recalling why the
extended symmetry notion of quantum double algebras is an optimal
tool when analyzing a wide variety of two dimensional physical
systems including quantum fluids, crystals and liquid crystals. The
power of this approach stems from the fact that one may characterize
both ordinary and topological modes as representations of a single
(generally non-Abelian) Hopf symmetry. In principle a full
classification of defect mediated as well as ordinary symmetry
breaking patterns and subsequent confinement phenomena can be given.
The formalism applies equally well to systems exhibiting global,
local, internal and/or external (i.e. spatial) symmetries. The
subtle differences in interpretation for the various situations are
pointed out. We show that the Hopf symmetry breaking formalism
reproduces the known results for ordinary (electric) condensates,
and we derive formulae for defect (magnetic) condensates which also
involve the phenomenon of symmetry restoration. These results are
applied in two papers which will be published in parallel
\cite{mbnematic:2006,bmmelting:2006}.

\end{abstract}

\end{frontmatter}

\newpage
\tableofcontents
\newpage
\section{Introduction and motivations}
\label{intro} In recent years it has become clear that in the
setting of two-dimensional (quantum) physics, the notion of quantum
double algebras or quantum groups plays an important role. One of
the main reasons is that these extended symmetry concepts allow for
the treatment of topological and ordinary quantum numbers on equal
footing. This means that the representation theory of the underlying
- hidden - Hopf algebra $\Acal$ labels both the topological defects
and the ordinary excitations. This is also the case in non-Abelian
situations where the mutual dependence of these dual quantum numbers
would otherwise be untractable. Moreover, as the Hopf algebras
involved turn out to be quasi-triangular they are naturally endowed
with a $R$-matrix which describes the topological interactions
between the various excitations in the medium of interest
\cite{Bais:1991pe,dwpb1995}. This theory has found interesting
applications in the domain of Quantum Hall liquids
\cite{Moore:1991ks,Read:1998ed,Nayak:1995kx,Slingerland:2001ea,Fendley:2005yy},
exotic phases in crystals and liquid crystals \cite{mbnematic:2006},
as well as 2-dimensional gravity \cite{Bais:1998yn}. It also appears
to furnish the appropriate language in the field of topological
quantum computation
\cite{Kitaev:1997wr,Preskill:1997ds,Freedman:2000rc,Dennis:2001nw}.

Once this (hidden) extended symmetry was identified, the authors of
\cite{Bais:2002pb,Bais:2002ny} studied the breaking the Hopf
symmetry by assuming the formation of condensates, respectively of
ordinary (which we call \emph{electric}), defect (\emph{magnetic}),
or mixed (\emph{dyonic}) type. As was to be expected, the ordinary
condensates reproduce the conventional theory of symmetry breaking,
though the analysis of confinement of topological degrees of
freedom, using the braid group, is not standard.  In
\cite{Bais:2002pb,Bais:2002ny} it was shown that when considering
Hopf symmetry breaking, the usual formalism of symmetry breaking had
to be extended with significant novel ingredients. One assumes the
condensate to be represented by a fixed vector in some nontrivial
representation of the Hopf algebra $\Acal$. This leads to the
definition of an intermediate algebra $\Tcal$ as the suitably
defined stabilizer subalgebra of the condensate. The complication
that arises at this level is that certain representations of $\Tcal$
may braid nontrivially with the condensate, which in turn means that
the condensate cannot be single valued around a particle belonging
to such a representation. If this happens to be the case, it implies
that such particles (representations) necessarily are confined. The
effective low energy theory of the non-confined degrees of freedom
is then characterized by yet a smaller algebra called $\Ucal$. So
the breaking of Hopf symmetries involves three algebras: the
unbroken algebra $\Acal$, the intermediate algebra $\Tcal$, and the
unconfined algebra $\Ucal$.

In this paper we show that the assumptions made about the
structure of the intermediate residual symmetry algebra $\Tcal$ in
\cite{Bais:2002pb,Bais:2002ny} can be relaxed. This leads to a new
definition of the residual symmetry algebra $\Rcal$ which contains
$\Tcal$ as a subalgebra, which means that the residual phase may have a richer spectrum.
We will discuss the new criterion in some detail and point out its importance.

We conclude by deriving general formulae for $\Rcal$ and $\Ucal$ for
the case of electric and defect condensates, in phases whose
corresponding Hopf algebra is what we call a \emph{modified quantum
double} $F(H_m)\times\C H_{el}$, where $H_m$ and $H_{el}$ are finite
groups. $H_m$ is the \emph{magnetic group}, i.e. the defect group.
$H_{el}$ is the \emph{electric group}, or the residual symmetry
group$^1$. We know what to expect for electric condensates, and for
that case our method reproduces the known results. The problem of
defect condensates is more interesting, and provided the main
motivation for this work. In this paper we focus on the basic
structure and the more formal aspects of the symmetry breaking
analysis and we mainly present general results, involving both
confinement and liberation phenomena. In two separate papers we give
detailed applications of these results: one paper on defect mediated
melting \cite{bmmelting:2006} (which discusses also the phenomenon
of symmetry restoration), and another on the classification of
defect condensates in non-abelian nematic crystals
\cite{mbnematic:2006}.

The starting point in this paper will be situations where a
continuous internal or external (space) symmetry $G$ is broken to a
possibly non-abelian, discrete subgroup $H$. This may happen through
some Higgs mechanism in which case we speak of \emph{discrete gauge
theories} \cite{Bais:1980fm}, or by forming some liquid crystal for
example. Topological (line) defects can then be labelled by the
elements $h$ of the discrete residual symmetry group$^2$ $H$.
Familiar examples of such defects are dislocations, disclinations,
vortices, flux-tubes etc. Such a theory has a (hidden) Hopf symmetry
corresponding to the so called \emph{quantum double} $D(H)$ of $H$.
This Hopf symmetry is an extension of the usual group symmetry$^3$.
The elements ${a}$ of $D(H)$ are denoted by $a=(f \times h)$, where
$f$ is some function on the group $H$, and $h$ some element of the
group algebra $\mathbb{C}H$ of $H$ (later on we will omit the
$\times$ sign). A basis for the space of functions $F(H)$ on a
discrete group $H$ is a set of delta functions which project on each
group element. We denote these projectors by $P_g$, these are just
delta functions $\delta_g$ ($g\in H$) defined by:
$\{P_g(g')\equiv\delta_{g}(g') = \delta_{g,g'} \;|\; \forall g' \in
H$ \}. The element $f$ is basically a measurement operator
(projecting on a subspace), while $h$ involves a symmetry operation.
In other words, if we denote the elements of the group $H$ by $g_i$,
where $i$ labels the group elements, we can write
\begin{eqnarray}
&& f = \sum_i \lambda_i P_{g_i} \nonumber\\
&& h = \sum_i \mu_i g_i, \nonumber
\end{eqnarray}
where the $\lambda_i$ and $\mu_i$ are complex numbers. This
follows from the fact that the $P_{g_i}$ span $F(H)$, and the
$g_i$ span the group algebra of $H$, which we denote by $\C H$. An
alternate notation for $D(H)$ is $F(H)\times\C H$, which shows
that it is a combination of $F(H)$ and $\C H$.

 The multiplication of two elements of $D(H)$ is defined by
\begin{equation}
(f_1\times h_1) (f_2\times h_2)(x) = f_1(x)f_2(h_1^{-1} x h_1)
\otimes h_1h_2  \;\;\;\; x\in H
\end{equation}
Note that the multiplication on the group part is just the ordinary
group multiplication, but the multiplication of the functions is not
just pointwise but twisted by a conjugation. This is a nontrivial
feature which implies that the product is not a simple tensor
product. There are two structures on a Hopf algebra $\Acal$ of
special interest to our purposes (for an introduction to Hopf
algebras, see for example \cite{Majid}). One is the so called
\textit{counit} $\varepsilon$ which turns out to be of particular
relevance in the present context. $\varepsilon$ is an algebra
morphism from the Hopf algebra to $\C$, defined through $
\varepsilon(f\times h) = f(e)$. The other structure we wish to
mention is the \textit{antipode} $S$ which is a kind of inverse
needed to introduce conjugate or antiparticle representations. $S$
is an algebra antimorphism from $\Acal$ to $\Acal$, which satisfies
$S(P_g \times h) \equiv P_{h^{-1}g^{-1}h} \times h^{-1}$. A quantum
double is further endowed with an invertible element $R \in \Acal
\otimes \Acal$ which implements the braiding of representations. It
encodes the topological interactions and in particular the possible
exotic ``quantum statistics properties'' of the excitations in the
system.

The sectors of the theory, or the physical excitations for that
matter, can be labelled by the irreducible representations of
$D(H)$. These are denoted as $\Pi^A_\alpha$, where the label $A$
denotes the conjugacy class $C_A \subset H$ to which the topological
(magnetic) charge of the excitation belongs, whereas $\alpha$
denotes a representation of the centralizer group $N_A$ of a chosen
preferred element $h_A$ of $C_A$. $\alpha$ fixes the ``ordinary'' or
electric charge. Clearly, $\Pi^A_\alpha$ describes in general a
mixed electric and magnetic, usually called \emph{dyonic},
excitation. Pure defects or ordinary excitations correspond to the
special cases where one of the two labels becomes trivial. We denote
the carrier space on which $\Pi^A_{\alpha}$ acts as $V_\alpha^A$.
The Hopf algebra structure requires a well defined comultiplication
which ensures the existence of well defined fusion or tensor product
rules for the representations of the algebra.

The quantum double appears when the defect group $H_m$, which we
call the magnetic group, and the residual symmetry group $H_{el}$,
which we call the electric group, are the same group $H$. This is
often the case, because the magnetic group is equal to
$H_m=\Pi_1(G/H_{el})=H_{el}$, when $G$ is connected and simply
connected. However, there are cases where $H_m$ and $H_{el}$ are not
equal. For example, if we want to distinguish between chiral and
achiral phases, we have to include inversions (or reflections). The
inclusion of inversions does not alter the defect group (because the
inversions live in a part of $G$ that isn't connected to the
identity), but it can alter the electric group. We are thus led to
the structure of a modified quantum double $F(H_m)\times\C H_{el}$,
which is a special case of what is called a bicrossproduct$^4$
 of the Hopf algebras $F(H_m)$ and $\C H_{el}$. The
structures in $F(H_m)\times\C H_{el}$ are very similar to those in
$D(H)$, we give them in appendix \ref{app:Hopf}. The crucial point
is that the electric group acts on the magnetic group, namely a
residual symmetry transformation $h\in H_{el}$ transforms the
defects. We denote the action of $h\in H_{el}$ on $g\in H_m$ by
$h\cdot g$. In the case of $D(H)$, the action is simply conjugation:
$h\cdot g = h g h^{-1}$. To obtain the structures of $D(H)$ from
those of $F(H_m)\times\C H_{el}$, one basically replaces all
occurrences of $h\cdot g$ by $hgh^{-1}$.

\section{Breaking, braiding  and confinement}

In the previous section we mentioned that a Hopf symmetry captures
the fusion and braiding properties of excitations of spontaneously
broken phases. In particular we mentioned the modified quantum
double $F(H_m)\times\C H_{el}$, which is relevant for a phase with
magnetic group $H_m$ and electric group $H_{el}$. In this section
we will show that this Hopf symmetry description allows for a
systematic investigation of phase transitions induced by the
condensation of some excitation. Since electric, magnetic and
dyonic modes are treated on equal footing, we will unify the study
of phase transitions induced by the formation of condensates of
all three types of modes.

The use of Hopf symmetries to analyze phase transitions was
pioneered by Bais, Schroers and Slingerland
\cite{Bais:2002pb,Bais:2002ny}. In this article we stay close to
their work. An important departure however, is the definition of
"residual symmetry operators". Based on physical arguments we put
forward a new definition, which, though similar, is more general
than the previous one.

\subsection{Breaking}

We want to study the ordered phases arising if a condensate forms,
corresponding to a non-vanishing vacuum expectation value of some
vector which we denote as $|\phi_0>$ in some representation of the
Hopf algebra. This vacuum vector breaks the Hopf-symmetry $\Acal$,
and just as in the conventional cases of spontaneous symmetry
breaking we have to analyze what the residual symmetry algebra
$\Tcal$ is.

The natural criterion to determine the residual symmetry that was
proposed in \cite{Bais:2002pb,Bais:2002ny} is
\begin{equation}
\Pi_{\phi}(a)\mid \phi_0 \rangle = \varepsilon(a)\mid \phi_0
\rangle \qquad\forall \quad a \in \Tcal, \label{eq:criterion1}
\end{equation}
where $\Pi_{\phi}$ is the representation in which $\mid \phi_0
\rangle$ lives. The motivation for this criterion is that an
operator $a$ is a residual symmetry operator if it acts on $\mid
\phi_0 \rangle$ as the trivial representation $\eps$.

This condition was subsequently analyzed in
\cite{Bais:2002pb,Bais:2002ny}, where it was argued that one must
impose further restrictions on $\Tcal$. Namely, it was imposed
that $\Tcal$ be the {\it maximal Hopf subalgebra} of $\Acal$ that
satisfies the condition (\ref{eq:criterion1}). This restriction
was made under the quite natural physical assumption that one
should be able to fuse representations of $\Tcal$.

We have taken a closer look at this restriction, and the point we
make in this paper is that $\Tcal$ need \emph{not} be a Hopf
algebra. It turns out that there is no physical reason to require
$\Tcal$ to be Hopf, and in fact lifting that requirement opens up
a number of very interesting new possibilities that appear to be
essential in describing realistic physical situations. Rather than
talking about a single Hopf-algebra $\Tcal$ we have to distinguish
two algebras (not necessarily Hopf) $\Rcal$ and $\Lcal$. One can
choose either one and we choose to define the breaking with
$\Rcal$. $\Rcal$ is defined as the set of operators in $\Acal$
that satisfy
\begin{equation}
(1\tensor \Pi_{\phi})\Delta(a)(1\tensor |\phi_0>) = a\tensor
|\phi_0>. \label{criterion2}
\end{equation}
We will explain the physical motivation for this criterion later on.
Stated in words: the residual symmetry operators belonging to
$\Rcal$ are operators that cannot distinguish whether a given
particle has fused with the condensate $|\phi_0>$ or not. Therefore
these operators are so to say ``blind'' to the condensate, and the
measurements they are related to are not affected by the presence of
the condensate. We note for now that $\Rcal$ need not be a Hopf
algebra, but that it contains the Hopf algebra $\Tcal$ as a
subalgebra. Still, all its elements $a$ satisfy
$\Pi_{\phi}(a)|\phi_0>=\eps(a)|\phi_0>$, thus the operators in
$\Rcal$ also act as the vacuum representation on $|\phi_0>$.

\begin{figure}[ht]
\begin{center}
\includegraphics[scale=0.6]{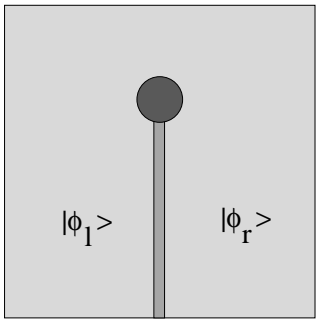}
\end{center}
\caption{{\small  A confined particle in the broken vacuum with a
domain wall attached. The state $|\phi_r>$ of the vacuum on the
right differs from the state $|\phi_l>$ of the vacuum on the left.}}
\label{phileftright}
\end{figure}

The important difference with the previous analysis of
\cite{Bais:2002pb,Bais:2002ny}, is that, as we have lifted the Hopf
algebra requirement on $\Tcal$, we will in general no longer have a
well defined tensor product for the representations of $\Rcal$. At
first sight this seems problematic from a physical point of view,
but the opposite turns out to be the case, the ambiguity that arises
reflects the physical situation perfectly well. Imagine that we
bring some localized excitation into the medium, moving it in from
the left, then it may well be that the vacuum is not single valued
when ``moved'' around this excitation. If such is the case the
excitation will be \emph{confined}, and we obtain a situation as
depicted in Figure \ref{phileftright}. The confined particle is
connected to a domain wall. The condensate to the right of the wall
is in the state $|\phi_r>$, and to the left it is in the state
$|\phi_l>$. Thus the question arises where the condensate takes on
the original value $|\phi_0>$. When we say the condensate is in the
state $|\phi_0>$, and we want to study confined excitations, we must
specify whether the state of the condensate is $|\phi_0>$ to the far
left or the far right of the plane, away from any excitations. We
cannot a priori impose that the state of the condensate to the far
left and the far right is the same, because then we wouldn't allow
for confined excitations. Choosing to set $|\phi_r>=|\phi_0>$, i.e.
setting the state of the condensate to the far right, gives $\Rcal$
as intermediate symmetry algebra. Choosing $|\phi_l>=|\phi_0>$ gives
us a different intermediate algebra $\Lcal$. As mentioned before
we'll work with $\Rcal$.

\begin{figure}[t]
\begin{center}
\includegraphics[scale=0.6]{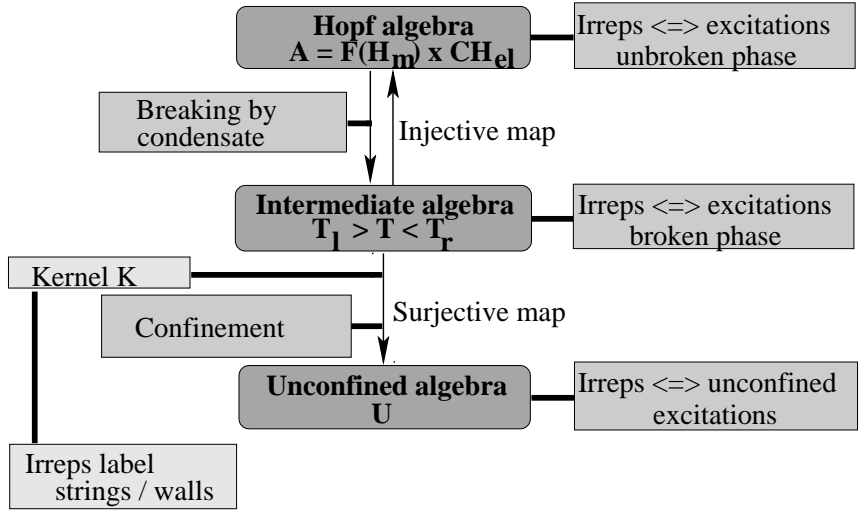}
\end{center}
\caption{{\small  A schematic of the quantum symmetry breaking
formalism. One has to distinguish three levels: The unbroken Hopf
algebra $D(H)$, the intermediate algebra $\Rcal$, and the effective
unconfined algebra $\Ucal$. This scheme should be compared with the
simple scheme of ordinary symmetry breaking where one only has to
distinguish between the unbroken  group $G$ and some residual
symmetry group $H \subset G$.}} \label{quantumbreaking}
\end{figure}

One may now restrict oneself to the unconfined representations and
obtain that they form the representations of yet another algebra,
the unconfined (Hopf) algebra $\Ucal$. Putting it all together we
have arrived at the picture in figure \ref{quantumbreaking} for the
generic symmetry breaking scheme.
The mathematical structure is appealing and exactly reflects the
subtlety of having to deal with confined particle
representations.\\
One may show that the maximal Hopf-subalgebra $\Tcal$ satisfying
the criterion (\ref{eq:criterion1}) is contained in the
intersection of $\Rcal$ and $\Lcal$:
\begin{equation}\label{t-relation}
  \Tcal\subseteq \Rcal \cap \Lcal.
\end{equation}
It is also possible to demonstrate that \emph{if} $\Rcal$ is a
Hopf algebra then we have necessarily the relation $\Tcal = \Rcal
= \Lcal$. We will show these and other results later.

In the following sections we motivate the Hopf symmetry breaking
formalism by way of examples, and we fully analyze electric and
defect condensates starting with phases whose Hopf symmetry is a
modified quantum double $F(H_m)\times\C H_{el}$. The breaking by
electric condensates proceeds along familiar lines (i.e.
Landau's theory of phase transitions). However, our arguments are
more complete also in these situations, because within our
formalism we arrive at a residual symmetry algebra $\Rcal$ where
all defects are still present, and we then find $\Ucal$ by
analyzing which defects are confined, and removing them from the
spectrum.

The fact that we reproduce the theory of electric condensates is
encouraging. More interesting nontrivial results occur when we
analyze defect condensates. Applications of the results of this
paper to defect condensates in general classes of nematic liquid
crystals \cite{mbnematic:2006}, and defect mediated melting
\cite{bmmelting:2006}, will be published elsewhere.

Excitations in the broken phase, which has $\Rcal$ as its symmetry
algebra, should form representations of $\Rcal$. We must now take
a close look at these excitations: some of them turn out to be
attached to a domain wall, i.e. their presence would require a
half-line singularity in the condensate, and such a wall costs a
finite amount of energy per unit length (as we will show later in
a simple example). To see this, we must first discuss braiding. It
is also useful to discuss the issue of braiding, to point out
characteristic differences between phases in which local and
global symmetries are broken.

\subsection{Braiding} Braiding addresses the question of
what happens to a multi-particle state when one particle is
adiabatically transported around another. To answer this question
one is led to introducing a braid operator, whose properties we
briefly recall. We focus our discussion on phases with a
$F(H_m)\times\C H_{el}$ Hopf symmetry. We start with some
heuristic arguments first involving the braiding of a defect and
an ordinary excitation, then the braiding of two defects, and
finally that of two ordinary excitations. We will assume
throughout that $H_{el}=H_m=H$ is discrete.

Consider the braiding of an electric mode with a defect in the gauge
theory case, because the global case is more subtle. This question
goes back to the very definition (or measurement) of the defect
\cite{Propitius}. If we have a discrete gauge theory with a defect
$g$, then $g$ is the ``topological charge'' of the defect, and it is
defined as the path-ordered exponential of the gauge field along a
path around the defect:
$$g=P e^{i\oint \vec{A}\cdot d\vec{x}}.$$
This phase factor basically corresponds to the change of the
wave function $|v>$ (which takes a value in some irreducible
representation of the gauge group $H$) of some particle coupled to
the gauge field when it is parallel transported around
the defect. The outcome is
\begin{equation}
\label{eq:pathordered} |v> \mapsto |v'>=P e^{i\oint \vec{A}\cdot
d \vec{x}}|v> = g\cdot|v>.
\end{equation}
So $|v'>$ is defined through the condition of a vanishing
covariant derivative: $D_i |v'>=0$.

Thus \emph{as an electric mode is transported around the defect, it
is acted on by the topological charge $g$ defining the defect}. We
can choose a gauge such that the field $v$ is constant everywhere
except along a line going ``up'' from the origin. Then half-braiding
already gives $v\mapsto g\cdot v$, because underneath the defect $v$
is constant, i.e. $v$ braids trivially with the defect if it passes
below the defect. This just shows that this distinction is gauge
dependent, because the operator itself is. Indeed, even the path
ordered exponential in ({\ref{eq:pathordered}) is not gauge
invariant in the nonabelian case, however the conjugacy class to
which it belongs is. In a locally invariant theory any physical
outcome can only depend on some gauge invariant expression involving
this path ordered exponential.

Let us now discuss the braiding of two defects
\cite{Bais:1980fm,dwpb1995}.  If we carry a defect $h$
counterclockwise around a defect $g$  then $h$ gets conjugated by
$g$ and becomes $ghg^{-1}$, while $g$ gets conjugated to $hgh^{-1}$.
We can encode this behavior by defining a \emph{braid operator}
$\braid$. If $|g>$ lives in $V^A$ and $|h>$ in $V^B$, then $\braid$
is a map from $V^A\tensor V^B$ to $V^B\tensor V^A$ whose action is
defined by (in the local theory this involves adopting some suitable
gauge fixing):
\begin{equation}
\braid\cdot |g> \otimes |h> = |ghg^{-1}> \otimes |g>
\label{eq:braidmagmag}
\end{equation}
The braid operator encodes the braiding properties of the defects.
Note that it braids the defect to the right halfway around the
other defect, it implements a \emph{half-braiding} or an
interchange. To achieve a full braiding, we have to apply the
\emph{monodromy} operator which equals $\braid^2$.\\
The equation for the braiding of defects $|g>$ and $|h>$ we have
just discussed applies equally well to the cases of global and
local symmetry breaking.

Finally, it is  clear that electric modes braid trivially with each
other$^5$:
\begin{equation}
\braid |v_1> \otimes |v_2> = |v_2> \otimes |v_1>.
\label{eq:braidelel}
\end{equation}

Before turning to more formal aspects of braiding let us comment on
the case of global symmetry. For example in a crystal the defects
are defined using a Burgers vector (giving a displacement after
circumventing a \emph{dislocation}) or a Frank vector (specifying a
rotation being the deficit angle after one carries a vector around a
\emph{disclination}). The first has to do with translations and the
other with discrete rotations, both are elements of the discrete
symmetry group of the lattice. The story is thus very similar to the
local case. The essence is that the local lattice frame is changed
while circumventing the defect (one speaks of ``frame dragging'')
and that will obviously affect the propagation of local degrees of
freedom, whether these are ordinary modes or other defects.

There is also a ``continuum approach'' to lattice defects where
these are considered as singularities in the curvature and torsion
of some metric space. In other words, the defects affect the space
around them, which will leave its traces if one is to bring a
particle around the defect. This idea has been used in crystals, and
the resulting geometry is of the Riemann-Cartan type$^6$ where the
disclinations and dislocations can be considered as singular nodal
lines of curvature and torsion respectively. For an introduction,
see \cite{Kleinert2} and \cite{Katanaev}. Clearly this geometrical
approach to defects assumes some suitable continuum limit to be
taken, which turns the theory into an $ISO(3)$ gauge theory (just
like gravity is an $SO(3,1)$ gauge theory).

Describing phases with global symmetries in terms of gauge fields
leads to the analog of the Aharonov-Bohm effect in global phases.
This has been studied in many phases, such as superfluid helium
\cite{Khazan}, crystals \cite{Furtado:2000}, and uniaxial nematic
liquid crystals \cite{McGraw} (neglecting diffusion).

So far it is clear that for the global theory, the outcome of
braiding is basically the same as in the local case. The frame
dragging however is locally measurable and cannot be changed by
local gauge transformations. This means that the global theory may
admit additional (nonivariant) observables.

We have seen that in basically all situations the action of the
braid operator on a two-particle state consisting of a defect and
an ordinary (electric) excitation is
\begin{eqnarray}
&&\braid\cdot (|g>\tensor|v>) = (\alpha(g)|v>)\tensor|g> \\
&&\braid\cdot (|v>\tensor|g>) = |g>\tensor|v>
\label{eq:braidelmag}
\end{eqnarray}

The beauty of many Hopf algebras $\Acal$ - and the quantum double is
one of them - is that they are \emph{quasitriangular}, which means
that they are naturally endowed with a \emph{universal R-matrix}
denoted by $R$. $R$ is an element of $\Acal\tensor\Acal$. It encodes
the braiding of states of the irreps of $\Acal$: to braid two
states, $|\phi_1>$ in $\Pi_1$ and $|\phi_2>$ in $\Pi_2$, act with
$R$ on $|\phi_1>\tensor|\phi_2>$, and then apply the flip operator
$\tau$. This gives the action of the braid operator $\braid$:
\begin{equation}
\braid (|\phi_1>\tensor |\phi_2>) = \tau\circ(\Pi_1\tens \Pi_2)
\circ R \circ |\phi_1> \tens |\phi_2>.
\end{equation}
The operator $\tau$ trivially interchanges any two vectors
$|\phi_1>$ and $|\phi_2>$ around:
\begin{equation}
\tau (|\phi_1>\tensor|\phi_2>) = |\phi_2>\tensor|\phi_1>.
\end{equation}

The braid (or rather monodromy) operator shows up in invariant
physical quantities. For example in the gauge theory case, the
expression for the cross section for the nonabelian analogue of
Ahoronov-Bohm scattering involving the monodromy operator was
given by E. Verlinde \cite{Verlinde}:
\begin{equation}
\frac{d\sigma}{d\theta}=\frac{1}{4\pi p
sin^2\te/2}[1-Re<\psi_{in}|\braid^2|\psi_{in}>]
\end{equation}
where $|\psi_{in}>$ is the initial internal wavefunction of the
whole system, and $p$ the incoming momentum. For example, for the
case of an electric mode $|v>$ to the left of a vortex $|g>$, we
have $|\psi_{in}>=|v>\tensor|g>$. If the braiding is trivial, i.e.
$\braid^2|\psi_{in}> = |\psi_{in}> $, then there is no scattering.

The question whether this expression for the cross section is also
valid in the case of global symmetries, has not been answered
conclusively, see for example \cite{March-Russell:1991az}. As far
as the analysis of conceivable ways to  break Hopf symmetries, the
general theory  applies to both situations, but in the actual
application there is a precise mathematical difference between
phases with local or global symmetries.

\subsection{Confinement}

We have already indicated that excitations which braid nontrivially
with the vacuum state cannot be strictly \emph{local} excitations,
because the condensate cannot be single valued around them. Such an
excitation can only be excited at the price of creating a physical
halfline singularity attached to it.  One has to create a
configuration of a domain wall ending at a defect and therefore we
say that such a defect is confined.  The fate - such as confinement
- of  defects in phase transitions where the topology changes was
discussed in relation with a particular exact sequence of homotopy
mappings in \cite{Bais:1980nn}. We will see that the condensate and
hence the vacuum vector takes a different value to the left and to
the right of the domain wall. And since the domain wall emanating
from the excitation costs a finite amount of energy per unit length,
we have to conclude that such an excitation will be linearly
\emph{confined}. A finite energy configuration requires another
confined excitation on the other end of the wall. We will call a
configuration consisting of a number of confined constituents
connected by a set of walls, a \emph{hadronic composite}, in analogy
with the hadrons in QCD, being the unconfined composites of confined
quarks (see Figure \ref{fig:confinement}). The excitations that
braid trivially with the condensate are clearly \emph{unconfined},
and they can propagate as isolated particles.

To illustrate some of these aspects in more detail, we take a look
at the ubiquitous XY-model in two dimensions, and reinterpret the
phase transition to the ordered state as an example where
\emph{confinement} plays a role.

\begin{figure}[t]
  \begin{center}
      \includegraphics[height=4cm]{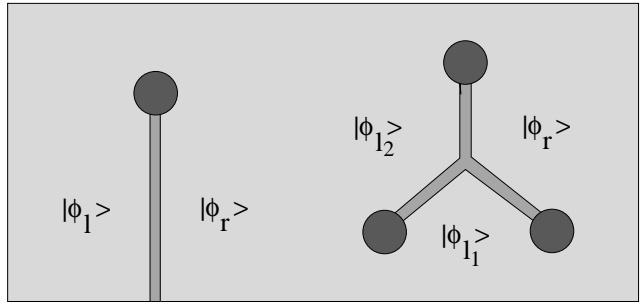}
      \caption{{\small A confined particle (left) and a hadronic composite (right).
    Confinement of particles in the ordered phase, and hadronic composites. The dark lines are
singularities in the condensate and cost a finite amount of energy
per unit length.}}
    \label{fig:confinement}
  \end{center}
\end{figure}

Let $\te(\vec{r})$ be the angular variable of the XY model, and
$\phi=|\phi|e^{i\te}$ the order parameter. The equation that
$\te(\vec{r})$ has to satisfy in order to minimize the free energy
of the XY model is Laplace's equation in two dimensions:
\begin{equation}
\nabla^2\te = 0 \label{eq:laplace}.
\end{equation}
There are point singular solutions $\te_{n,\vec{r_0}}$ of
(\ref{eq:laplace}) that correspond to a defect of charge $n$
centered at $\vec{r_0}$. There are also other singular solutions of
(\ref{eq:laplace}), in which the singularity is a half-line. They
are labelled by a $\lambda\in\R$ and a vector $\vec{r_0}=(x_0,y_0)$
and the corresponding solution is given by
\begin{equation}
\te_{\lambda,\vec{r_0}}(x,y)=\eta \; arctan (\frac{x-x_0}{y-y_0}) =
\lambda \varphi,
\end{equation}
with $\varphi$ the polar angle.

Note that for $\lambda=n\in\Z$, there is indeed no line singularity.
For $\lambda\notin\Z$, there is a line starting at $\vec{r_0}$ and
going out to infinity, along which $\phi=v e^{i\te}$ is
discontinuous (we assume $|\phi|=v$ is constant in the ordered
phase). To see that there is a discontinuity, we follow a loop
around $\vec{r_0}$, and notice that as we go full circle $\te$ turns
by an angle $2\pi\lambda$. If $\lambda\notin\Z$, $\te$ does not
return to its original value as we finish travelling along our loop
(remember that $\te$ is defined modulo $2\pi$). Thus there is a half
line singularity in $\te$, which implies a half line singularity in
$\phi$. One easily checks that this line singularity carries a
finite amount of energy per unit length. Thus, if $\lambda\notin\Z$,
the free energy of the configuration $\te_{\lambda,\vec{r}_0}$
increases linearly with the system size. Such a wall with an end is
not a topological defect in the strict sense, since it does not
carry a $\Pi_1(G/H)$ charge, but its appearance can be understood
from analyzing the appropriate exact homotopy sequence
\cite{Bais:1980nn} . We will call it a vortex with non-integer
charge $\lambda$, and conclude that this vortex is \emph{confined}.
It is attached to a half-line singularity which corresponds to a
\emph{domain wall}, because the "line" bounded by the defect is of
one dimension lower than the dimension of the space. The usual
definition implies that an excitation is confined if its energy
increases linearly with the system size.
\begin{figure}[t]
  \begin{center}
  {(a) \includegraphics[height=4cm]{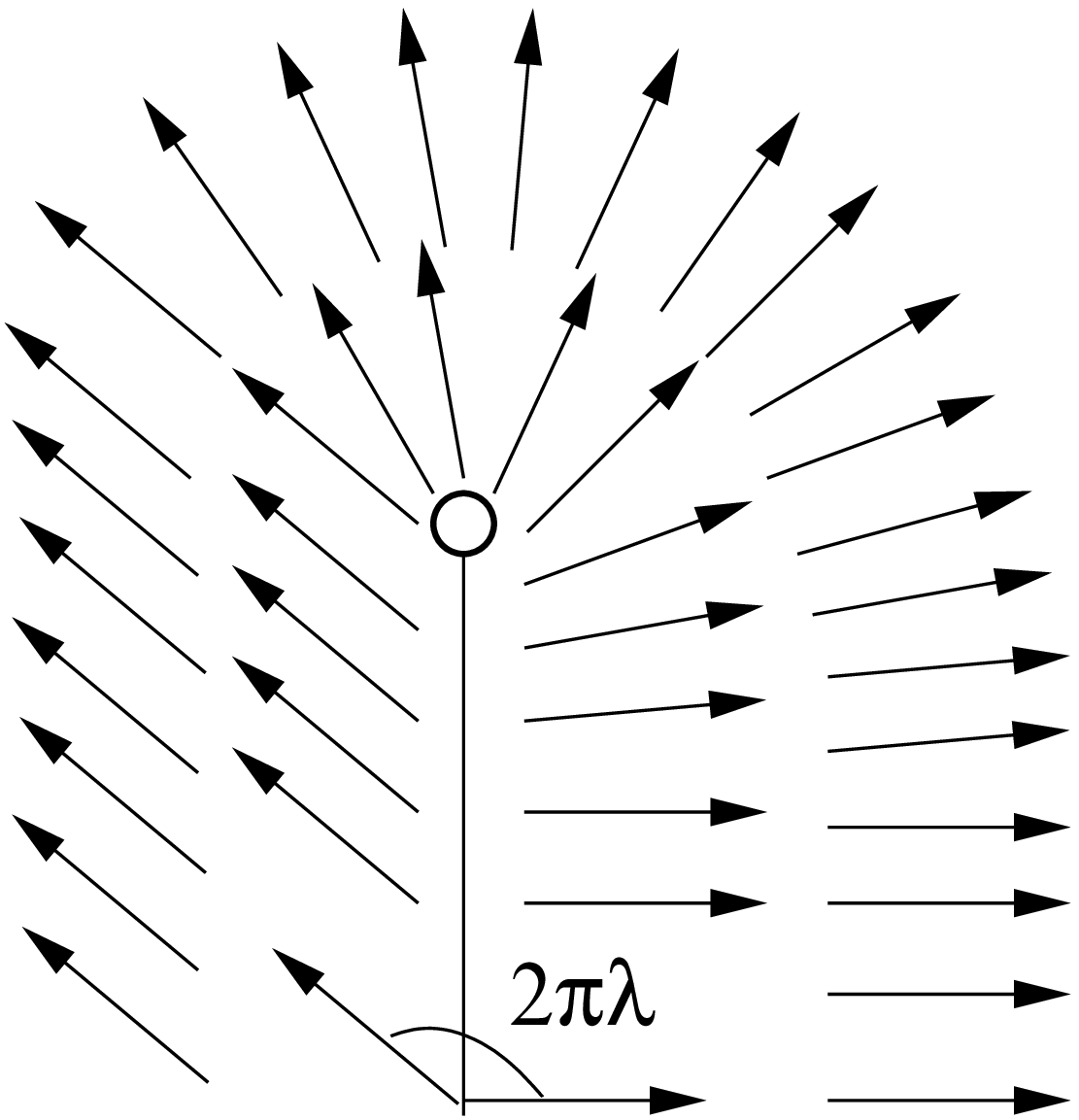}}
  {(b) \includegraphics[height=4cm]{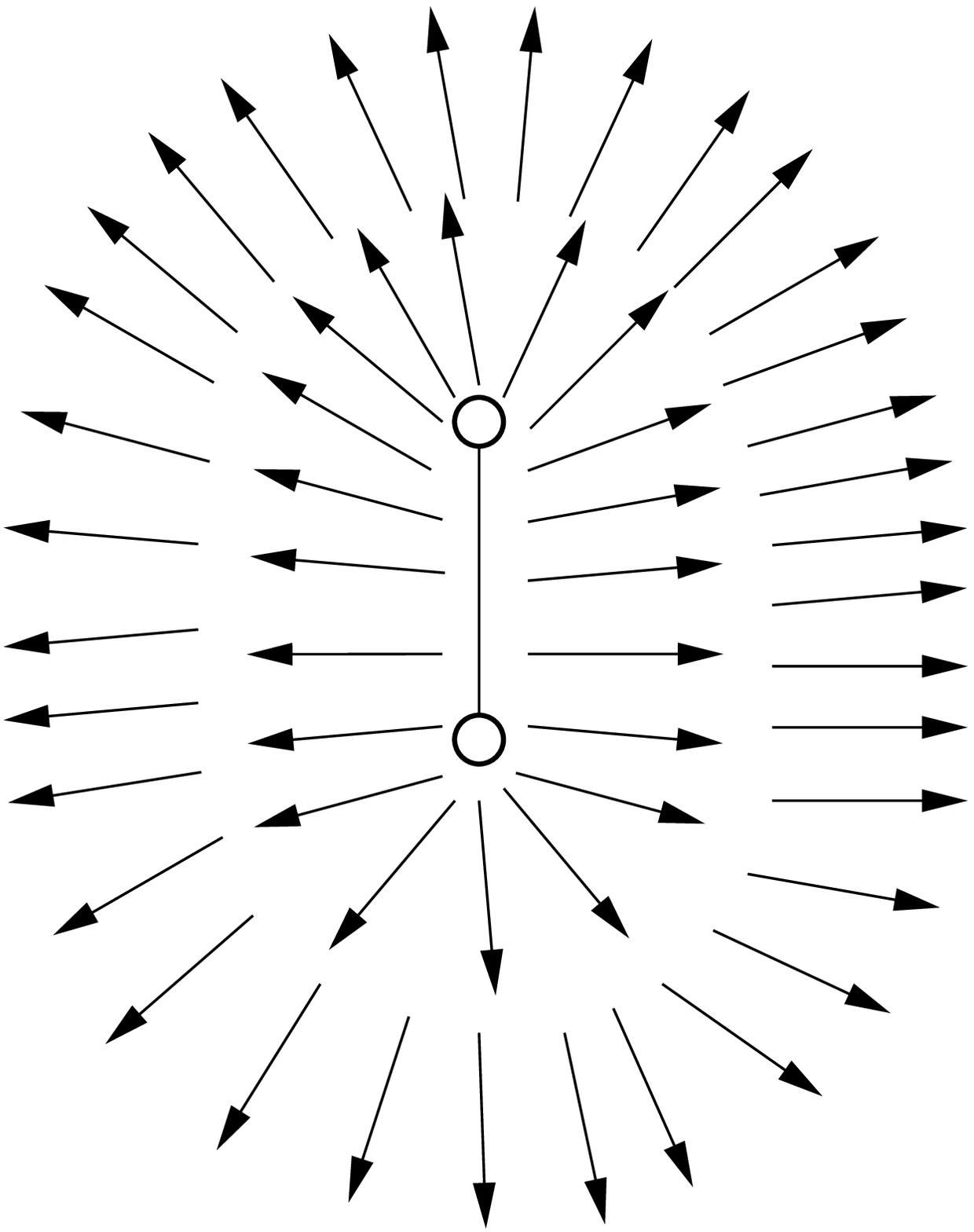}}
    \caption{{\small  Confinement of vortices of non-integer charge
$\lambda$ in the ordered phase of the XY-model. The arrows point in
the direction of the local value of $\te$. In (a) we see a confined
charge $\lambda$ vortex in the XY model. A half-line singularity
starts at the core of the defect (schematically indicated by the
small circle) and goes out to infinity. It carries a finite amount
of energy per unit length. (b) shows an unconfined composite of two
confined charge $\frac{1}{2}$ vortices. The charge $\frac{1}{2}$
vortices are attached by a singular line. The overall configuration
has charge 1.}} \label{fig:conf_XY}
  \end{center}
\end{figure}

After symmetry breaking, the vortices of
non-integer charge are confined. It is conceivable that there are
noninteger charges in this phase, connected together by half-line
singularities, such that the overall charge is integer. See Figure
\ref{fig:conf_XY} for an example of a hadronic composite of two
charge $\frac{1}{2}$ vortices.

As mentioned, the crucial characteristic of confined excitations
is that the condensate takes on a different value to the left and
right of the excitation. We started by condensing the order
parameter field, such that it took on the value $|\phi_0>$
everywhere. Now we have an excitation such that the order
parameter field takes the value $|\phil>$ to the left, and
$|\phir>$ to the right of the half-line singularity connected to
the excitation. To unambiguously define the vacuum state
$|\phi_0>$, we must make a choice how we treat the excitations.
This we do by specifying that excitations, both confined and
unconfined, enter the system from the left.  Thus, we set
$|\phi_0>=|\phir>$, and as a confined excitation comes in from the
left the condensate to the left of the excitation takes on the
value $|\phil>\neq|\phi_0>$. If an unconfined excitation comes in
from the left, then $|\phil> = |\phir> = |\phi_0>$.

Imagine a particle in a state $|v>$ coming in from the left. We want
to know whether $|v>$ is confined of not, i.e. we need to know if
$|\phil>=|\phir>$, in which case $|v>$ is unconfined. This is where
braiding comes in: $|\phil>$ is the outcome of half-braiding
$|\phir>$ counterclockwise around $|v>$. Thus, to find out whether
$|v>$ is confined, we braid the condensate around the excitation,
using the braid operator $\braid$. For $|v>$ to be unconfined, the
condensate $|\phir>$ has to braid trivially around $|v>$, both
clockwise and counterclockwise. We must check that both braidings
are trivial, because the half-line singularity could run along the
positive or the negative x-axis in fig. \ref{fig:conf_XY}. This
leads to the following test of whether an excitation is confined or
unconfined:
\begin{eqnarray}
& \textrm{$|v>$ is unconfined} &\nonumber\\
& \iff& \nonumber
\end{eqnarray}
\begin{displaymath}
\left\{ \begin{array}{c}
    \braid\cdot (|v>\tensor|\phir>) = |\phir>\tensor|v>\\
    \braid^{-1}\cdot (|v>\tensor|\phir>) = |\phir>\tensor|v>.\end{array}\right. \nonumber
\end{displaymath}

Let us apply this criterion to the vortices in the XY model, to see
which ones are unconfined. Denote a vortex of charge $\lambda$ by
$|\lambda>$. We set $|\phir>=v$. Then $|\phir>$ gets frame dragged
as we braid it counterclockwise around $|\lambda>$, and picks up a
phase factor $e^{i2\pi\lambda}$. Clockwise braiding is trivial,
since we've chosen the convention that the clockwise braiding of an
electric mode to the right of a defect is trivial$^7$. Thus
\begin{eqnarray}
&& \braid\cdot (|\lambda> \tensor |\phir>) = e^{i2\pi\lambda} |\phir> \tensor |\lambda>\nonumber\\
&& \braid^{-1}\cdot (|\lambda>\tensor|\phir>) = |\phir>\tensor
|\lambda>,\nonumber
\end{eqnarray}
and a vortex is unconfined $\iff \lambda\in\Z$. Thus the defects are
precisely the unconfined vortices.

In the high temperature phase, the ``electric'' excitations are
irreps of the electric group $U(1)$. If we also consider projective
irreps, then the electric excitations are irreps of the universal
covering group $\widetilde{U(1)}=\R$ of $U(1)$. The irreps are
denoted by $\rho_r$, they are labelled by an $r\in \R$. All these
irreps braid trivially with the condensate $|\phi_r>=v$, because the
condensate is in a state $|\phir>$ in an electric irrep, and
electric irreps braid trivially with each other. Thus in the ordered
phase all the electric excitations are unconfined. The situation is
different from the situation where defects condense as is discussed
in \cite{bmmelting:2006}.

So indeed the phase transition from the high temperature to the
ordered phase in the XY-model can in this sense be described as a
confining transition.

\section{Hopf symmetry breaking: the formalism}

In this section we will first study symmetry breaking for a phase
described by a general Hopf algebra $\Acal$. Then we will
specialize to the cases where $\Acal$ is a modified quantum
double: \mbox{$\Acal=F(H_m)\times \C H_{el}$}. The quantum double
$D(H)$ is a special case of the modified quantum double, with
$H_m=H_{el}=H$.

\subsection{The criterion for symmetry breaking} We consider a
physical system whose excitations (or particles)  are labeled by
irreps of a quasitriangular Hopf algebra $\Acal$. We say that a
condensate forms in a state $|\phir>$ of an irrep $\Pi_{\phi}$ of
$\Acal$, so that our ground state is a state filled with the
particles in the state $|\phir>$. We now have to define the
\emph{residual symmetry algebra} $\Rcal$, which is a symmetry
algebra with the property that the excitations of the new ground
state form irreps of $\Rcal$.

The residual symmetry algebra $\Rcal$ consists of the operators that
are `well defined with respect to the condensate'. Before we explain
what that means, let us first consider the case of a system whose
symmetry is an ordinary group $G$. $G$ is broken spontaneously to
$H$ by a condensate in a state $|\phir>$ of an irrep of $G$. $H$ is
the stabilizer of $|\phir>$, i.e. the subset of symmetry
transformations in $G$ that leave the condensate invariant. And
because they leave the condensate invariant we can implement these
transformations on excitations of the condensate. Thus they act on
the excitations, and  since by definition they also commute with the
Hamiltonian, they transform low energy excitations into excitations
of same energy, which can therefore be organized in irreps of H .

Now consider a particle in state $|v>$ of an irrep of the original
symmetry $G$. If we fuse $|v>$ with $|\phir>$, and act on the
outcome $|v>|\phir>$ of this fusion with an $h\in H$, the result
is the same as when we act on $|v>$ with $h$ first, and then fuse
the outcome with $|\phir>$. This follows from
\begin{equation}
h\cdot (|v>\tensor|\phir>) =
(h\cdot|v>)\tensor(h\cdot|\phir>)=(h\cdot|v>)\tensor |\phir>,
\end{equation}
since $h\cdot|\phir>=|\phir>$. Thus we can also define the
residual symmetry group $H$ to be the set of transformations that
\emph{are insensitive to fusion with the condensate}. Whether a
particle $|v>$ in an irrep of $G$ fuses with the condensate or
not, the action of $H$ is the same. Thus we define \emph{residual
symmetry operators} to be operators that are not affected by
fusion with the condensate. When the symmetry is spontaneously
broken from $G$ to $H$, the residual symmetry operators are the
elements of $H$.

This definition of residual symmetry operator can be carried over
to the case of a condensate $|\phir>$ in an irrep of a Hopf
symmetry $\Acal$. The residual symmetry operators are the
operators that are not affected by fusion with the condensate. The
residual symmetry operators form a subalgebra of $\Acal$, which we
call the \emph{residual symmetry algebra} $\Rcal$. There is a
subtlety in the definition of $\Rcal$: an operator $a\in \Acal$ is
an element of $\Rcal$ if its action on any particle $|v>$ in any
irrep of $\Acal$ is the same whether $|v>$ has fused with the
condensate $|\phir>$ or not. But we must specify whether $|v>$
fuses with $|\phir>$ from the left or the right! Namely, in the
systems we are considering $|v>\tensor |\phir>$ and $|\phir>
\tensor |v>$ are not (necessarily) the same state, because of the
possibility of nontrivial braiding. Thus we have to fix a
convention. This convention is set by our earlier choice of having
all particles come in from the left. Remember that we had to make
this choice because some excitations of the condensate may be
confined. Thus we define $\Rcal$ as follows: $a\in\Rcal\iff$ for
any particle $|v>$ in any irrep of $\Acal$, we have
\begin{equation}
a\cdot(|v>\tensor|\phir>) = (a\cdot|v>)\tensor|\phir>.
\end{equation}

Since $|v>$ and $|\phir>$ are states of particles of $\Acal$,
their fusion is set by the coproduct $\Delta$ of $\Acal$:
\begin{equation}
a\cdot (|v>\tensor|\phir>) =
(\Pi\tensor\Pi_{\phi})\circ\Delta(a)(|v>\tensor|\phir>),
\end{equation}
where $|v>$ is in the irrep $\Pi$ of $\Acal$, and $|\phir>$ in the
irrep $\Pi_{\phi}$ of $\Acal$. Since this equation has to hold for
all vectors $|v>$ in all irreps $\Pi$ of $\Acal$, it is equivalent
to$^8$
\begin{equation}
(1 \otimes \Pi_{\phi}) \Delta(a) (1 \otimes |\phir>) = a \otimes
|\phir>. \label{eq:Tresid}
\end{equation}
$\Rcal$ consists of all operators that satisfy this criterion. We
will now prove that $\Rcal$ is a subalgebra of $\Acal$. We also
prove two more properties of $\Rcal$ which will play a role later
on.
\begin{lem}
The elements of a finite dimensional Hopf algebra $\mathcal{A}$
that satisfy
\begin{equation}
(1 \otimes \Pi_{\phi}) \Delta(a) (1 \otimes |\phir>) = a \otimes
|\phir> \nonumber
\end{equation}
form a subalgebra $\Rcal$ of $\Acal$ that satisfies:
\begin{enumerate}
\item $\Delta(\Rcal) \subset \mathcal{A} \otimes \Rcal$ \item The
elements of $\Rcal$ leave tensor products of the vacuum invariant.
\end{enumerate}
\end{lem}
\begin{pf}

That $\Rcal$ is an algebra follows from the fact that $\Delta$ is
an algebra morphism.

\begin{enumerate}
\item Using the definition of $\Rcal$ and the coassociativity of
the coproduct,
\begin{eqnarray}
&& (1 \tensor 1 \tensor \Pi_{\phi}) (1 \tensor \Delta) \Delta(a)
(1 \tensor |\phir>) = \nonumber \\
&&= (1 \tensor 1 \tensor
\Pi_{\phi}) (\Delta \tensor 1) \Delta(a) (1\tensor |\phir>)= \nonumber \\
&& = (\Delta\tensor 1) (1\tensor\Pi_{\phi}) \Delta(a) (1\tensor
|\phir>) \nonumber \\
&& = (\Delta \tensor 1)(a\tensor |\phir>
=\sum_{(a)}a^{(1)} \tensor a^{(2)} \tensor |\phir> \nonumber \\
&& \Rightarrow \sum_{(a)} a^{(1)} \tensor (\Delta(a^{(2)}) \cdot(1
\tensor |\phir>)) = \sum_{(a)} a^{(1)}
\tensor a^{(2)} \tensor |\phir> \nonumber \\
&& \Rightarrow \Delta(a^{(2)})\cdot (1 \tensor |\phir>) = a^{(2)} \tensor \phi \nonumber \\
&& \Rightarrow a^{(2)} \in \Rcal \nonumber \\
&& \Rightarrow \Delta(a) \in \mathcal{A} \tensor \Rcal \nonumber
\end{eqnarray}
\item We just proved that $(1 \tensor \Pi_{\phi}) \Delta(a^{(2)})
(1\tensor |\phir>) = a^{(2)} \tensor |\phir>$. Using this, we get
\begin{eqnarray}
&& (1\tensor \Pi_{\phi} \tensor \Pi_{\phi}) (1 \tensor \Delta)
\Delta(a) (1 \tensor |\phir> \tensor
|\phir>)\nonumber\\
&&= \sum_{(a)} a^{(1)} \tensor ((\Pi_{\phi} \tensor 1) (1 \tensor
\Pi_{\phi}) \Delta(a^{(2)}) (1\tensor
|\phir>) (|\phir> \tensor 1)) \nonumber \\
&&= \sum_{(a)} a^{(1)} \tensor ((\Pi_{\phi} \tensor 1) (a^{(2)}
\tensor |\phir>) (|\phir> \tensor 1))
\nonumber\\
&&= ((1 \tensor \Pi_{\phi}) \Delta(a) (1\tensor |\phir>)) \tensor |\phir> \nonumber\\
&&= a \tensor |\phir> \tensor |\phir> \nonumber
\end{eqnarray}
\end{enumerate}
\end{pf}
We call $\Rcal$ the \emph{right residual symmetry algebra}.\\
What if we condense the sum of two vectors in different irreps,
$|\phi_1> + |\phi_2>$? According to (\ref{eq:Tresid}), $a\in\Acal$
is part of the right residual symmetry algebra $\Rcal$ of
$|\phi_1> + |\phi_2>$ if
\begin{eqnarray}
(1\tensor\Pi_{\phi_1}) (\Delta(a)) (1\tensor |\phi_1>) &+&
(1\tensor\Pi_{\phi_2}) (\Delta(a)) (1\tensor |\phi_2>)\nonumber \\
&=& a\tensor |\phi_1> +a \tensor |\phi_2>.
\end{eqnarray}
Since we are dealing with irreps, the only way to get an equality is
by equating the first terms on the left-hand side and right-hand
side of this equation, and the last terms, separately. So $\Rcal =
(\Rcal)_1 \cap (\Rcal)_2$, the intersection of the right residual
symmetry algebras of $|\phi_1>$ and $|\phi_2>$. Therefore, we need
only treat the condensation of vectors in one irrep, since we can
then take intersections of the right residual symmetry algebras of
condensates in different irreps to get the right residual symmetry
algebra for
any condensate. \\

We will now specialize to the case where $\Acal$ is a modified
quantum double: \[\mbox{$\Acal=F(H_m)\times\C H_{el}$},\] where
$H_m$ and $H_{el}$ are groups (see Appendix \ref{app:Hopf}).

The irreps of $\Acal=F(H_m)\times\C H_{el}$ are labelled by an
orbit $A$ in $H_m$ under the action of $H_{el}$, and an irrep of
the normalizer $N_A$ of a preferred element $g_A\in A$.

In Appendix \ref{app:Hopf} we show how to write elements of
$F(H_m)\times\C H_{el}$ as functions $f\in F(H_m\times H_{el})$.
In this notation the derivations to come are more elegant.
Consider symmetry breaking by a condensate $|\phir>$ in an irrep
$\Pi^A_{\alpha}$ of $\Acal$.

\begin{lem}
\label{lem:R} Take $f \in F(H_m\times H_{el})$. Then
\begin{eqnarray}
f \in \Rcal &\iff& f(x_1 (x\cdot g_A),y_1) |\phir(y_1^{-1}x)> =
f(x_1,y_1) |\phir(x)> \nonumber \\  &\forall& x_1 \in H_m, x, y_1
\in H_{el} \label{eq:Tresqd}
\end{eqnarray}
\end{lem}
\begin{pf}
We use the formulae in Appendix \ref{app:Hopf}.
\begin{eqnarray}
&&(id \otimes \Pi^A_{\alpha})\Delta(f)( 1 \otimes |\phir(x)>) = f \otimes |\phir(x)> \nonumber \\
\iff && 1 \otimes \sum_{z \in H_{el}} \Delta(f)(x_1,y_1;x\cdot
g_A,z)|\phir(z^{-1}x)>=f(x_1,y_1) \otimes
|\phir(x)>\nonumber\\
\iff && 1 \otimes \sum_{z \in H_{el}} f(x_1 (x\cdot g_A),y_1)
\delta_{y_1}(z) |\phir(z^{-1}x)>= f(x_1,y_1)
\otimes |\phir(x)> \nonumber\\
\iff && 1 \otimes f(x_1 (x\cdot g_A),y_1) |\phir(y_1^{-1}x)> =
f(x_1,y_1) \otimes |\phir(x)> \nonumber
\end{eqnarray}
\end{pf}
$\Rcal$ was obtained by condensing $|\phir>$ to the right of our
system. If we choose to condense $|\phir>$ to the left, we get
another residual symmetry algebra:
\begin{equation}
\Lcal = \{a\in\Acal:(\Pi_{\phi} \otimes id) \Delta(a)(|\phir>
\otimes 1) = |\phir> \otimes a\}.
\end{equation}
We call $\Lcal$ the \emph{left residual symmetry algebra}.

It is interesting to compare this criterion to the one in
\cite{Bais:2002ny}. There, the \emph{residual symmetry algebra} is
denoted by $\Tcal$, and is defined as the largest Hopf subalgebra
of $\Acal$ whose elements satisfy
\begin{equation}
\label{eq:Tcrit}
 a\cdot |\phir> = \eps(a) |\phir>.
 \end{equation}
The motivation for this criterion is: The residual symmetry
operators act on the condensate like the vacuum irrep $\eps$ does.

The important difference between $\Tcal$ and $\Rcal$ is that we
don't require $\Rcal$ to be a Hopf algebra. In particular, we
don't expect the residual symmetry to have a coproduct. In fact
there is a physical reason why we don't need to have a coproduct
for $\Rcal$. We have chosen the condensate to be $|\phir>$ on the
right. If we bring in a confined excitation from the left, the
condensate will be in a state $|\phil>$ to the left of this
excitation. Now consider a second particle coming in from the
left, it sees the condensate $|\phil>$, and it is therefore an
excitation associated with the residual symmetry of $|\phil>$,
which needn't be equal to the residual symmetry of $|\phir>$!
Thus, we have to keep track of the \emph{ordering} of the
particles, i.e. it is crucial to know in which order we brought in
the particles from the left. We will see that - not surprisingly -
this translates into the absence of a coproduct in $\Rcal$, namely
we can't simply fuse irreps of $\Rcal$. Before we can discuss this
in more detail, we need to take a closer look at the structure of
the residual symmetry algebras.

\subsection{Relationship between $\Tcal$, $\Rcal$ and $\Lcal$}

We will now establish a number of interesting connections between
$\Tcal$, $\Rcal$ and $\Lcal$. All the operators in $\Rcal$ and
$\Lcal$ satisfy (\ref{eq:Tcrit}):
\begin{equation}
a\cdot |\phir> = \eps(a) |\phir>.\nonumber
\end{equation}
Thus the operators of $\Rcal$ and $\Lcal$ act on the condensate
like the vacuum irrep $\eps$ does, just as the operators of
$\Tcal$ do. At the same time, we have the inclusions
\begin{equation}
\Tcal\subset \Rcal \quad \Tcal\subset\Lcal.
\end{equation}
The left and right residual symmetry algebras contains $\Tcal$. We
prove these statements in the following lemma.
\begin{lem}
\label{lem:Res} $\Rcal, \Lcal$ and $\Tcal$ satisfy the following:
\begin{enumerate}
\item All elements of $\Rcal$ and of $\Lcal$ satisfy
(\ref{eq:Tcrit}):
\begin{equation}
a\cdot |\phir> = \eps(a) |\phir> \nonumber
\end{equation}
\item $\Tcal \subset \Rcal\cap\Lcal$
\end{enumerate}
\end{lem}
\begin{pf}
\begin{enumerate}
\item $a\in\Rcal$ implies:
\begin{eqnarray}
&&(id \otimes \Pi_{\phi}) \Delta(a) 1 \otimes |\phir> = a \otimes |\phir> \nonumber \\
\Rightarrow && (\eps \tensor 1) (1 \otimes \Pi_{\phi}) \Delta(a) 1
\otimes |\phir> = \eps(a) \otimes |\phir>
\nonumber \\
\Rightarrow && (1 \otimes \Pi_{\phi}) (\eps \tensor 1)\Delta(a) 1
\otimes |\phir> = \eps(a) \otimes |\phir>
\nonumber \\
\Rightarrow && 1 \tensor \Pi_{\phi}(a) |\phir> = \eps(a) \otimes |\phir> \nonumber \\
\Rightarrow && \Pi_{\phi}(a) |\phir> = \eps(a) |\phir>,\nonumber
\end{eqnarray}
where we used one the axioms of a Hopf algebra: $(\eps\tensor
id)\Delta(a)=1\tensor a$. This proves the claim for $\Rcal$. The
proof for $\Lcal$ is analogous. \item $\Tcal$ is a Hopf algebra, so
\begin{eqnarray}
&a\in&\Tcal \nonumber \\
&\Rightarrow&
\Delta(a)=\sum_{(a)}a^{(1)}\tensor a^{(2)}
\in \Tcal \tensor\Tcal \nonumber \\
&\Rightarrow& a^{(2)}\in
\Tcal\Rightarrow a^{(2)}\cdot |\phir>=\eps(a^{(2)}) |\phir> \nonumber\\
&\Rightarrow& (id\tensor\Pi_{\phi})\Delta(a)(1\tensor |\phir>)=(id
\tensor \eps) \Delta(a) (1\tensor |\phir>) =
a \tensor |\phir> \nonumber\\
&\Rightarrow& a\in\Rcal.\nonumber
\end{eqnarray}
This proves $\Tcal\subset\Rcal$. The proof that $\Tcal\subset\Lcal$
is analogous.
\end{enumerate}
\end{pf}
$\Rcal$ and $\Lcal$ are not necessarily Hopf algebras, while $\Tcal$
is a Hopf algebra by definition. It turns out that $\Rcal$ is a Hopf
algebra $\iff \Rcal=\Tcal$! Similarly, $\Lcal$ is a Hopf algebra
$\iff \Lcal=\Tcal$. Also, $\Rcal=\Tcal\iff \Rcal = \Lcal$. Thus
$\Rcal$ and $\Lcal$ are interesting extensions of $\Tcal$: if they
are equal to each other, then they are equal to $\Tcal$. Thus, the
difference between $\Rcal$ and $\Lcal$ is a measure of the departure
of $\Rcal$ (and $\Lcal$) from being a Hopf algebra.

We need one assumption about $\Acal$ to prove these propositions:
the antipode $S$ of $\Acal$ must satisfy $S^2=id$. Modified quantum
doubles, for example, satisfy this property.

First we prove a little lemma.
\begin{lem}
\label{lem:antipode} If the antipode $S$ of $\Acal$ satisfies $S^2
= id$, then $S(\Lcal)=\Rcal$ and $S(\Rcal)=\Lcal$.
\end{lem}
\begin{pf}
According to ($\ref{eq:antipode}$), and using $S^2 = id$:
\begin{eqnarray}
&&\Delta^{op} \circ S = (S\tensor S)\circ \Delta\nonumber\\
&&\Rightarrow \Delta^{op} = (S\tensor S)\circ \Delta\circ S
\nonumber
\end{eqnarray}
Say $a \in \Lcal$, then using the last equation
\begin{eqnarray}
&&(\Pi_{\phi} \otimes 1) \Delta(a) |\phir> \otimes 1 = |\phir> \otimes a \nonumber \\
&&\Rightarrow (1 \otimes \Pi_{\phi}) \Delta^{op}(a) 1 \otimes |\phir> = a \otimes |\phir> \nonumber \\
&&\Rightarrow (1 \otimes \Pi_{\phi}) (S\tensor S)\Delta(S(a)) 1
\otimes |\phir> = (a\tensor 1) (1 \otimes
|\phir>) \nonumber \\
&& \textrm{Apply $S\tensor S$ to left the and right, and use $S^2=id$ and $S(1)=1$ to get} \nonumber \\
&&\Rightarrow (1 \otimes \Pi_{\phi}) \Delta(S(a)) 1 \otimes |\phir> = S(a) \otimes |\phir> \nonumber \\
&&\Rightarrow (1 \otimes \Pi_{\phi}) \Delta(S(a)) 1 \otimes |\phir> = S(a) \otimes |\phir> \nonumber \\
&&\Rightarrow S(a) \in \Rcal \nonumber
\end{eqnarray}
So $S(\Lcal) \subseteq \Rcal$. Similarly we can prove that
$S(\Rcal)\subseteq\Lcal$. Since $S$ is invertible, we have
$dim(S(\Rcal))=dim(\Rcal)$ and $dim(S(\Lcal))=dim(\Lcal)$, where
$dim$ is the dimension as a vector space. Therefore, using
$S(\Lcal) \subseteq \Rcal$ and $S(\Rcal) \subseteq \Lcal$, we get
\begin{equation}
dim(S(\Lcal)) \leq dim(\Rcal) = dim (S(\Rcal)) \leq
dim(\Lcal)=dim(S(\Lcal)).
\end{equation}
Thus $dim(\Rcal)=dim(S(\Lcal))$. Since $S(\Lcal)\subseteq\Rcal$,
we must have $S(\Lcal)=\Rcal$. Applying $S$ to both side of this
equation, we get $\Lcal=S(\Rcal)$.
\end{pf}
This lemma states that the antipode $S$ brings us from $\Rcal$ to
$\Lcal$, and back. In appendix \ref{app:Hopf} we show how $S$ is used to construct the
antiparticle or conjugate irrep of a given irrep. Thus, going from
$\Rcal$ to $\Lcal$ is tantamount to replacing all particles by
their antiparticles!

Using lemma \ref{lem:antipode}, we can prove all our propositions
about the relationships between $\Tcal$, $\Rcal$ and $\Lcal$.
\begin{prop}
\label{prop:res} For $\Acal$ an n-dimensional Hopf algebra whose
antipode $S$ satisfies $S^2=id$, we have
$$(1) \Rcal = \Lcal \iff (2) \textrm{ $\Rcal$ is a Hopf algebra} \iff (3)\Rcal = \Tcal \iff (4) \Lcal
\subseteq \Rcal$$
\end{prop}
\begin{pf}
\begin{itemize}
\item $(1) \Rightarrow (2)$ \\
We assume $\Rcal=\Lcal$. Take $a \in \Rcal = \Lcal$. To prove that
$\Rcal$ is a Hopf subalgebra of $\Acal$, we need to prove three
things:
\begin{equation}
1 \in \Rcal,\quad \Delta(a) \in \Rcal \tensor \Rcal, \quad
S(\Rcal) \subset \Rcal.
\end{equation}
The first demand is trivial, because $\Delta(1) = 1 \tensor 1$, so
that
\begin{equation}
(1\tensor\Pi_{\phi})\Delta(1)(1 \tensor |\phir>) = 1\tensor
\Pi_{\phi}(1) |\phir> = 1 \tensor |\phir> \Rightarrow 1 \in \Rcal.
\end{equation}
For the second demand: since $a\in\Rcal=\Lcal$, we have $\Delta(a)
\in \Lcal \tensor \Acal = \Rcal \tensor \Acal$, and $\Delta(a) \in
\Acal \tensor \Rcal$. Choose a basis $\{r_i\}_{1\leq i \leq k}$ of
$\Rcal$, and a basis $\{a_j\}_{k+1 \leq j \leq n }$ of
$\Rcal^{\perp}$. Then
\begin{eqnarray}
&& \Delta(a) \in \Rcal \tensor \Acal \Rightarrow \Delta(a) = \sum r_i \tensor a'_i \nonumber \\
&&\textrm{Write $a_i'$ out in terms of the bases $\{r_i\}$ of
$\Rcal$ and $\{a_j\}$ of
$\Rcal^{\perp}$:}\nonumber \\
&& a'_i = B_{ij} r_j + C_{ik} a_k \quad B_{ij}, C_{ik} \in \C \nonumber \\
&& \Rightarrow \Delta(a) = \sum r_i \tensor B_{ij} r_j + \sum r_i \tensor C_{ik} a_k \nonumber\\
&& \Delta(a) \in \Acal \tensor \Rcal \Rightarrow C_{ik} a_k = 0 \nonumber \\
&& \Rightarrow \Delta(a) = \sum r_i \tensor B_{ij} r_j \in \Rcal
\tensor \Rcal \nonumber
\end{eqnarray}
Now for $S(\Rcal) \subseteq \Rcal$. Using lemma \ref{lem:antipode},
$S(\Rcal)=\Lcal=\Rcal$.

Thus $\Rcal$ is a Hopf subalgebra of $\Acal$.
\item $(2) \Rightarrow (3)$ \\
According to lemma \ref{lem:Res}, we have $\Tcal \in \Rcal$, and
all $a\in\Rcal$ satisfy $\Pi_{\phi}(a) |\phir> = \epsilon(a)
|\phir>$. $\Tcal$ was defined as the largest Hopf subalgebra of
$\Acal$ whose elements satisfy $\Pi_{\phi}(a) |\phir> =
\epsilon(a) |\phir>$, so if $\Rcal$ is a Hopf algebra we must have
$\Rcal=\Tcal$.
\item $(3) \Rightarrow (1)$ \\
We already proved that $\Tcal \subseteq \Lcal$. Thus if we assume
$\Tcal=\Rcal$ we have $\Rcal \subseteq \Lcal$.

Since $\Rcal=\Tcal$, $\Rcal$ is a Hopf algebra. Thus $S(\Rcal)
\subseteq \Rcal$. From lemma \ref{lem:antipode}, we know that
$S(\Rcal) = \Lcal$. Thus $\Lcal \subseteq \Rcal$.

Done: $\Rcal = \Lcal$.
\item $(1)\Rightarrow (4)$\\
Obvious.
\item $(4)\Rightarrow (1)$\\
Lemma \ref{lem:antipode} taught us that $\Rcal=S(\Lcal)$ and
$S(\Rcal)=\Lcal$. Apply $S$ to the left and right of $\Lcal
\subseteq \Rcal$ to obtain $\Rcal \subseteq \Lcal$. Done:
$\Rcal=\Lcal$.
\end{itemize}
\end{pf}

\subsection{$\Rcal$ and $\Lcal$: Hopf or not?}

We are interested in finding out which condensates yield a right
residual symmetry algebra $\Rcal$ that is a Hopf algebra. From
proposition \ref{prop:res} we know that $\Rcal$ is a Hopf algebra
$\iff$ $\Lcal$ is the same Hopf algebra.

As a rather general case, consider a phase with a modified quantum
double as its Hopf symmetry: $\Acal=F(H_m)\times\C H_{el}$. We can
write elements of $\Acal$ as functions $f\in F(H_m\times H_{el})$
(see appendix \ref{app:Hopf}). Now condense $|\phir>$ in an irrep
$\Pi^A_{\alpha}$ of $\Acal$.  We saw in lemma \ref{lem:R} that a
function $f\in F(H_m \times H_{el})$ is an element of $\Rcal$ if it
satisfies
\begin{eqnarray}
f(x_1 (x\cdot g_A),y_1) |\phir(y_1^{-1}x)> = f(x_1,y_1) |\phir(x)>
\quad &\forall& x_1 \in G,\nonumber \\ &x,& y_1 \in H_{el}
\nonumber
\end{eqnarray}
where $g_A$ is the preferred element of $A$.

Analogously to the derivation of lemma \ref{lem:R}, we can prove
that the functions $f$ in $\Lcal$ are precisely those $f$ that
satisfy
\begin{eqnarray}
f((x\cdot g_A) x_1,y_1) |\phir(y_1^{-1}x)> = f(x_1,y_1) |\phir(x)>
\quad &\forall& x_1\in G, \nonumber \\ &\forall& x, y_1 \in
H_{el}. \label{eq:TLresqd}
\end{eqnarray}

Proposition \ref{prop:res}(4) tells us that proving that $\Rcal$
is a Hopf algebra is equivalent to proving that $\Lcal\subseteq
\Rcal$. Thus, to prove that $\Rcal$ is a Hopf algebra, we must
prove that if a function $f$ satisfies (\ref{eq:TLresqd}), it
automatically satisfies (\ref{eq:Tresqd}):
\begin{eqnarray}
&&\forall x_1\in H_m, \forall x, y_1\in H_{el}: \nonumber\\
&&f(x_1 x\cdot g_A,y_1) |\phir(y_1^{-1}x)>= f(x_1,y_1) |\phir(x)> \nonumber\\
\Rightarrow &&f( (x\cdot g_A) x_1, y_1) |\phir(y_1^{-1}x)> =
f(x_1,y_1) |\phir(x)> \label{eq:TrHopf}
\end{eqnarray}

This implication is automatically satisfied if $H_m$ is an abelian
group, because then $x\cdot g_A$ and $x_1$ commute. Thus, if the
magnetic group $H_m$ is abelian, $\Rcal$ is necessarily a Hopf
algebra.

Also, if $g_A$ is in the center of $H_m$, and is acted on
trivially by all of $H_{el}$, then (\ref{eq:TrHopf}) is satisfied.
Electric condensates are an example, since $g_A=e$ for electric
condensates. $e$ is in the center of $H_m$, and $H_{el}$ acts
trivially on $e$.

If the original phase is a $D(H)$ phase, then we have similar
results: $\Rcal$ is a Hopf algebra if $H$ is abelian, or if $g_A$
is in the center of $H$. We needn't demand that all of $H$ acts
trivially on $g_A$, since this immediately follows from $g_A$
being in the center of $H$.

\subsection{Requirement on the condensate $|\phir>$}

If the condensate is $|\phir>$, then our ground state is filled
with the particles in the state $|\phir>$. We know that if an
excitation of this ground state braids nontrivially with the
condensate, then it is connected to a domain wall which
costs a finite amount of energy per unit length. Thus, if
$|\phir>$ were to braid nontrivially with itself, it wouldn't make
sense to think of a condensate. Thus we require of our condensate
$|\phir>$ that it braid trivially with itself:
\begin{equation}
\braid \circ( |\phir> \tensor |\phir>) = |\phir> \tensor |\phir>.
\label{eq:TSB}
\end{equation}
Note that we are braiding indistinguishable particles, and if
$|\phir>$ has spin $s$, then the braiding picks up an extra phase
factor $e^{i2\pi s}$. The spin factor should be taken into account
when verifying the trivial self braiding condition.

Recently fermionic condensates have received considerable
theoretical and experimental attention\cite{Jin}, and we could
definitely treat those as well with our methods. We may then relax
the trivial self braiding condition exactly because identical
fermions don't braid trivially with each other: they pick up a
minus sign under half-braiding.

\subsection{Unconfined excitations and the algebra $\Ucal$}

\subsubsection{The conditions of trivial braiding}

Now that we've learned how to derive $\Rcal$, we want to study the
unconfined excitations of $\Rcal$. We found a criterion for
determining whether an excitation was confined: If the excitation
doesn't braid trivially with the condensate $|\phir>$, then it is
confined.

The condensate is in a state $|\phir>$ of an irrep $\Pi_{\phi}$ of
$\Acal$. Now consider an excitation of the ground state, sitting in
an irrep $\Omega$ of $\Rcal$. Since the universal R matrix
$R=\sum_{(R)}R^{(1)}\tensor R^{(2)}\in \Acal\tensor\Acal$, we cannot
simply act with $R$ on states in the tensor product representation
$\Pi_{\phi}\tensor\Omega$, because we can only act with elements of
$\Rcal$ on states of $\Omega$. We need a projection $P$ of $\Acal$
onto $\Rcal$, so that we can act with \mbox{$(id\tensor P) R$} on
states of $\Pi_{\phi}\tensor\Omega$. If $\Acal$ has an inner
product, then we can use this inner product to define the projection
P. Take an orthonormal basis $\{a_i\}$ of $\Rcal$, and an
orthonormal basis$^9$ $\{b_j\}$ of $\Rcal^{\perp}$. Together, the
$a_i$ and $b_j$ form an orthonormal basis of $\Acal$, such that the
$a_i$ span $\Rcal$, and for all $i$ and $j$ we have $^{10}$
\begin{equation}
(a_i,b_j) = 0,
\end{equation}
where $(a,b)$ denotes the inner product between $a$ and $b$. Now
take any $a\in\Acal$, and write
\begin{equation}
a = \sum_{(a_i)}(a_i,a) a + \sum_{(b_j)}(b_j,a)b_j \equiv Pa +
(1-P)a.
\end{equation}
Thus we have our projection: $P a = \sum_{(a_i)}(a_i,a) a \in
\Rcal$. It is known that this projection is in fact independent of
our choice of basis: given a vector space with an inner product,
the perpendicular projection onto a vector subspace is uniquely
defined.

Modified quantum doubles $F(H_m)\times\C H_{el}$ come equipped
with an inner product (\ref{eq:inner}):
\begin{equation}
(P_g h, P_{g'}h') = \delta_{g,g'}\delta_{h,h'}. \nonumber
\end{equation}
We use this inner product to define the projection operator
\mbox{$P:\Acal\rightarrow\Rcal$}, just as we discussed above. Now
we can define the braiding of a state $|v>$ of $\Omega$ with
$|\phir>$:
\begin{eqnarray}
&\textrm{Counterclockwise}&: \tau \circ (\Omega\otimes \Pi^A_{\alpha}) (P\tensor id) R (|v> \tensor |\phir>)\\
&\textrm{Clockwise}&: \tau\circ (\Omega\tensor \Pi^A_{\alpha})
(P\tensor id) R^{-1}_{21}(|v>\tensor |\phir>)
\end{eqnarray}
where in the second line we used: $\braid^{-1}=\tau\circ
R_{21}^{-1}$.

We want to find out which irreps $\Omega$ of $\Rcal$ braid
trivially with the condensate, thus the braiding has to be trivial
for all $|v>$ in the vector space on which $\Omega$ acts. We still
need a definition of "trivial braiding". A natural definition is:
$\Omega$ braids trivially with the condensate if it braids just
like the vacuum irrep $\eps$ does. This definition immediately
implies that the vacuum irrep $\eps$ is unconfined, since
obviously $\eps$ braids like $\eps$
 does. Thus the conditions for trivial braiding of an irrep $\Omega$ of $\Rcal$ with the condensate $|\phir>$
living in the irrep $\Pi^A_{\alpha}$ of $F(H_m)\times\C H_{el}$
become
\begin{equation}
\label{eq:braid1} (\Omega\otimes \Pi^A_{\alpha})(P\tensor
1)R(1\tensor |\phir>) = (\Omega(1) \epsilon \tensor
\Pi^A_{\alpha})(P\tensor 1)R(1\tensor |\phir>)
\end{equation}
\begin{eqnarray}
\label{eq:braid2} (\Omega\tensor \Pi^A_{\alpha}) (P\tensor id)
R^{-1}_{21} &&(1\tensor|\phir>) = \nonumber \\ && =(\Omega(1) \eps
\tensor \Pi^A_{\alpha}) (P\tensor id) R^{-1}_{21} (1\tensor
|\phir>).
\end{eqnarray}
$\Omega(1)$ is an $n\times n$ unit matrix, where $n$ is the
dimensionality of the irrep $\Omega$.

If these equations are satisfied, then we can replace
$1\tensor|\phir>$ in these equations by $|v>\tensor |\phir>$ for
any state $|v>$ of $\Omega$. Thus if $\Omega$ satisfies the
trivial braiding conditions (\ref{eq:braid1}) and
(\ref{eq:braid2}), then all the states $|v>$ of $\Omega$ braid
trivially with $|\phir>$.

An irrep $\Omega$ that satisfies these two equations is said to
braid trivially with the condensate. If an irrep doesn't braid
trivially with the condensate, it is a confined excitation,
attached to a physical string that goes out to infinity which
costs a finite amount of energy per unit length.

\subsubsection{The unconfined symmetry algebra $\Ucal$}

The trivial braiding equations (\ref{eq:braid1}) and
(\ref{eq:braid2}) divide the irreps of $\Rcal$ into confined and
unconfined irreps. We cannot simply take the tensor product of
irreps of $\Rcal$, since $\Rcal$ isn't a Hopf algebra. The reason
for the absence of a coproduct is the presence of confined
excitations. The condensate to the right of a confined excitation
takes on the value $|\phir>$, while it takes on a different value
$|\phil>$ to the left of the excitation. Thus particles coming in
from the left see a different condensate: they are excitations of
the residual symmetry algebra of $|\phil>$.

The situation is actually a little more complicated, because the
value $|\phil>$ of the condensate to the left of a state $|v>$ of
an irrep $\Omega$ of $\Rcal$ depends on the explicit state $|v>$
of $\Omega$. Thus $|\phil>$ is not unique for an irrep $\Omega$.

We will discuss how to deal with these issues later. For now, we
note that unconfined excitations do not suffer from such
complications, since the condensate takes on a constant value
around unconfined excitations. Thus we expect the fusion rules of
unconfined excitations to be associative, if we only consider
their composition with other unconfined excitations. There should
be a Hopf algebra $\Ucal$, called the \emph{unconfined symmetry
algebra}, whose irreps are precisely the unconfined irreps, and
whose fusion rules give the fusion channels of the unconfined
excitations into other unconfined excitations.

To obtain $\Ucal$, we first determine all unconfined irreps of
$\Rcal$. Then we take the intersection of the kernels of all
unconfined irreps$^{11}$, and denote it by $\Ical$. Finally, we
define the algebra
\begin{equation}
\Ucal = \Rcal/\Ical
\end{equation}
This is an algebra because $\Ical$ is an ideal (i.e. an invariant
subalgebra) of $\Rcal$. Its irreps are precisely the
unconfined irreps.

Our claim is that \emph{$\Ucal$ is a Hopf algebra}. Though we have
not attempted to prove this in full generality, we found it to be
true in all the cases we've worked out. In view of our discussion
earlier, it is physically natural that $\Ucal$ is a Hopf algebra,
while $\Rcal$ needn't be. However the mathematical proof if this
conjecture may not be so easy, and this definitely deserves further
study. Such a proof should make use of the trivial self braiding
condition (\ref{eq:TSB}) that we formulated for the condensate
$|\phi_0>$, because if this condition is dropped then we have found
cases where $\Ucal$ isn't a Hopf algebra.

\subsubsection{Trivial braiding for $F(H_m/B)\tensor\C N$}

In the next section, we will see that for the electric and defect
condensates in a phase with $F(H_m)\times\C H_{el}$ symmetry, the
residual symmetry algebra $\Rcal$ takes on a special form:
\begin{equation}
\Rcal = F(H_m/B)\times\C N, \label{eq:Rcalspec}
\end{equation}
where $B$ is a subgroup of $H_m$, and $N$ is a subgroup of
$H_{el}$ whose elements $n \in N$ satisfy
\begin{equation}
n\cdot B = B.
\end{equation}
This last equation tells us that the action of $n\in N$ on $H_m/B$
is well defined: $F(H_m)\times\C H_{el}$ is a modified quantum
double, so that the action of $h\in H_{el}$ on $H_m$ satisfies
(\ref{eq:modqdact}):
\begin{equation}
\forall g_1,g_2\in H_m: h\cdot (g_1g_2)=(h\cdot g_1)(h\cdot
g_2).\nonumber
\end{equation}
Since $N\subset H_{el}$, all $n\in N$ also satisfy this equation.
The action of $n$ on $H_m/B$ is given by:
\begin{equation}
n\cdot (g B) = (n\cdot g)(n\cdot B) = (n\cdot g) B.
\end{equation}
$\Rcal=F(H_m/B)\times\C N$ is a transformation group algebra, so
we can use the canonical theorem on the irreps of transformation
group algebras given in Appendix \ref{app:Hopf}. The irreps are
labelled by an orbit $T$ in $H_m/B$ under the action of $N$, and
an irrep $\tau$ of the normalizer $N_T$ of a preferred element
$g_T$ of $T$. We denote irreps of $F(H_m/B)\tensor\C N$ by
$\Omega^T_{\tau}$. The conditions (\ref{eq:braid1}) and
(\ref{eq:braid2}) for $\Omega^T_{\tau}$ to braid trivially with
$|\phir>$ in $\Pi^A_{\alpha}$ reduce to
\begin{equation}
\label{eq:Ucalone} \sum_{h\in B} |\phir(\Gamma(h (x_{\eta}\cdot
g_T^{-1})) x)> = \sum_{h\in B} |\phir(\Gamma(h) x)>
\end{equation}
\begin{eqnarray}
\label{eq:Ucaltwo} &|\phir(x)>\neq 0, \Gamma(x\cdot g_A) \in N
\Rightarrow &\tau_{ij}(x_{\eta}^{-1} \Gamma(x\cdot g_A^{-1})
x_{\zeta})1_{N_T}(x_{\eta}^{-1} \Gamma(x\cdot g_A^{-1}) x_{\zeta}) \nonumber\\
&&= \tau_{ij}(x_{\eta}^{-1} x_{\zeta}) 1_{N_T}(x_{\eta}^{-1}
x_{\zeta})
\end{eqnarray}
where the $x_{\eta}$ and $x_{\zeta}$ are chosen representatives of
left $N_T$ cosets of $H$.

The proof of these equations is rather lengthy. Some intermediate
steps are
\begin{itemize}
\item Proof of (\ref{eq:Ucalone})
\begin{eqnarray}
&& R = \sum_{h,g\in G}P_g e \tensor P_h\Gamma(g)\nonumber\\
&&P(P_h \Gamma(g))=1_N(\Gamma(g))\frac{1}{|B|} P_{hB}\Gamma(g) \nonumber\\
&&\sum_{g,h} \eps(P(P_ge))\Pi^A_{\alpha}(P_h\Gamma(g)) |\phir(x)>
=\sum_{g\in B} \frac{1}{|B|}
|\phir(\Gamma(g)x)> \nonumber \\
&& \Omega^T_{\tau}(P(P_h
g))^{i,j}_{\eta,\zeta}=\frac{1}{|B|}1_N(\Gamma(g))
\sum_{N_T}P_{hB}(x_{\eta} \cdot
g_T) \delta_g(x_{\eta}n x_{\zeta}^{-1})\beta_{i,j}(n)\nonumber\\
&&\sum_{h,g} \Omega^T_{\tau}(P(P_g e))^{i,j}_{\zeta,\eta} \Pi^A_{\alpha} (P_h \Gamma(g)) |\phir(x)> \nonumber\\
&&= \Omega^T_{\tau}(1)^{i,j}_{\zeta,\eta} \sum
\eps(P(P_ge))\Pi^A_{\alpha} (P_h\Gamma(g)) |\phir(x)>
\nonumber\\
&&\iff
\frac{1}{|B|}\tau_{i,j}(x_{\eta}^{-1}x_{\zeta})1_{N_T}(x_{\eta}^{-1}x_{\zeta})
\sum_{ h\in B}
|\phir(\Gamma(h x_{\eta}\cdot g_T^{-1})x)>\nonumber\\
&& = \frac{1}{|B|}
\tau_{i,j}(x_{\eta}^{-1}x_{\zeta})1_{N_T}(x_{\eta}^{-1}x_{\zeta})\sum_{h
\in B} |\phir(hx)>
\nonumber\\
&&\iff \sum_{h\in B} |\phir( \Gamma(h (x_{\eta}\cdot
g_T^{-1}))x)>=\sum_{h\in B} |\phir(hx)> \nonumber
\end{eqnarray}
\item Proof of (\ref{eq:Ucaltwo})
\begin{eqnarray}
&& \sum_{h,g} \eps(P(P_h \Gamma(g^{-1})))\Pi^A_{\alpha}(P_g e)
|\phir(x)> = 1_N(\Gamma(x\cdot g_A)) |\phir(x)>
\nonumber\\
&& \sum_{h,g} \Omega^T_{\tau}(P(P_h
\Gamma(g^{-1})))\Pi^A_{\alpha}(P_g e)\phi= \nonumber \\ && \qquad
= \Omega^T_{\tau}(1)\sum_{h,g}
\eps(P(P_h \Gamma(g^{-1})))\Pi^A_{\alpha}(P_g e) |\phir(x)> \nonumber \\
&&\iff 1_N(\Gamma(x\cdot g_A))\tau_{i,j}(x_{\eta}^{-1}\Gamma
(x\cdot g_A^{-1})x_{\zeta}) 1_{N_T} (
x_{\eta}^{-1} \Gamma(x\cdot g_A^{-1})x_{\zeta}) |\phir(x)>\nonumber\\
&&=1_N(\Gamma(x\cdot
g_A))1_{N_T}(x_{\eta}^{-1}x_{\zeta})\tau_{i,j}(x_{\eta}^{-1}x_{\zeta})
\phi(x) \nonumber
\\
&&\iff |\phir(x)> = 0 \textrm{ or } \Gamma(x\cdot g_A^{-1}) \notin N \nonumber\\
&&\textrm{or } \tau_{i,j}(x_{\eta}^{-1}\Gamma (x\cdot
g_A^{-1})x_{\zeta}) 1_{N_T}(x_{\eta}^{-1}\Gamma(x\cdot
g_A^{-1})x_{\zeta})=1_{N_T} (x_{\eta}^{-1}x_{\zeta}) \tau_{i,j}
(x_{\eta}^{-1}x_{\zeta}) \nonumber
\end{eqnarray}
\end{itemize}

When the modified quantum double is a quantum double $D(H)$, the
conditions for an irrep $\Omega^T_{\tau}$ of $\Rcal$ to braid
trivially with the condensate $|\phir>$ become
\begin{equation}
\label{eq:UcaloneD} \sum_{h\in B} |\phir(h x_{\eta}
g_T^{-1}x_{\eta}^{-1} x)> = \sum_{h\in B} |\phir(h x)>
\end{equation}
\begin{eqnarray}
\label{eq:UcaltwoD} &|\phir(x)> \neq 0, xg_Ax^{-1} \in N
\Rightarrow &\tau_{ij}(x_{\eta}^{-1} xg_A^{-1}x^{-1} x_{\zeta})
1_{N_T}(x_{\eta}^{-1} xg_A^{-1}x^{-1} x_{\zeta}) \nonumber\\
&&= \tau_{ij}(x_{\eta}^{-1} x_{\zeta}) 1_{N_T}(x_{\eta}^{-1}
x_{\zeta}).
\end{eqnarray}

We will now use these equations to work out $\Rcal$ and $\Ucal$ for
various condensates in modified quantum doubles. We start with
electric condensates, and show that the conventional theory of
electric condensates (Landau's theory) is reproduced. After that we
study a variety of defect condensates.
\begin{figure}[ht]
\centering
\includegraphics[width=8cm,keepaspectratio=true]{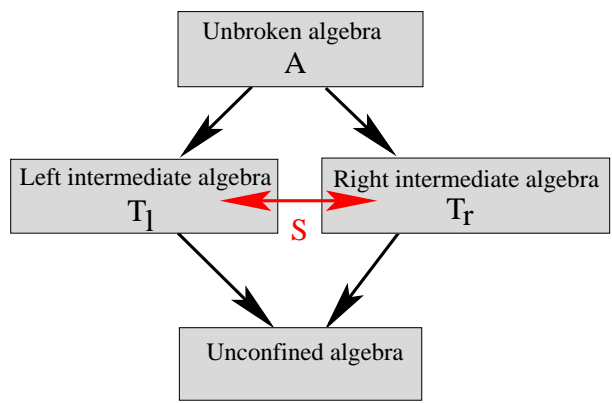}
\caption{{\small  The symmetry breaking scheme.}} \label{uniaxial}
\end{figure}
\section{Condensates in modified quantum doubles}
In the previous sections we formulated the general conditions which
determine the residual symmetries in the case where a Hopf symmetry
is spontaneously broken by the formation of a condensate in a state
$|\phir>$. We were led to consider an intermediate algebra $\Rcal$
(or $\Lcal$) and an unconfined algebra $\Ucal$ which would be
manifest in the low energy spectrum of the broken phase. In this
section we will present some rather general derivations of $\Rcal$
and $\Ucal$ for certain classes of vacua $|\phir>$.

We start with a brief discussion of "conventional" symmetry breaking
by an ordinary (electric) type condensate. We recover the well known
results with the added feature that the confinement of certain types
of defect follows from our formalism. This was already obtained in
\cite{Bais:2002ny}, but with a different definition of the residual
symmetry algebra.

In the next subsection we analyze the situations that arise if one
considers defect condensates. We point out that these can be of
various types leading to different low energy phenomena. We refrain
from discussing the case of mixed (or dyonic) condensates. These can
definitely be studied using our formalism, but they have to be dealt
with on a case to case basis, while for the case of ordinary and
defect condensates we were able to extract general formulae.

\subsection{Ordinary (electric) condensates}
Consider a phase described by a quantum double $D(H)$, and condense
a state $|\phir>$ of an electric irrep $\Pi^e_{\alpha}$ of $D(H)$.
Then equation (\ref{eq:Tresqd}) tells us that the functions $f$ in
$\Rcal$ satisfy
\begin{eqnarray}
&&f(x_1,y_1) \alpha(x^{-1}y_1) = f(x_1,y_1) \alpha(x^{-1}) \nonumber \\
&&f(x_1,y_1) \alpha(y_1) = f(x_1,y_1) \textbf{1} \nonumber \\
&& f(x_1,y_1)=0 \vee y_1 \in N_{\phir} \nonumber \\
\Rightarrow &&\Rcal = F(H) \otimes \mathbb{C} N_{\phir} \nonumber\\
\end{eqnarray}
where $N_{\phir}$ is the stabilizer of $|\phir>$, i.e. the set of
elements $h\in H$ that satisfy $\Pi^e_{\alpha}(h) |\phir> =
\alpha(h) |\phir> = |\phir>$. $\textbf{1}$ is the unit matrix.
Since the condensate is purely electric, the magnetic group is
unbroken, thus all of $F(H)$ is present in $\Rcal$. $\Rcal$ is a
Hopf algebra in this case, which implies that the tensor product
of irreps is well defined, and associative.

Some defects are confined. The trivial braiding conditions
(\ref{eq:UcaloneD}) and (\ref{eq:UcaltwoD}) tell us that only the
defects $g\in N_{\phir}$ are unconfined. Thus the unconfined
symmetry algebra is
\begin{equation}
\Ucal = D(N_{\phir}).
\end{equation}

One of the consequences of electric symmetry breaking is the
lifting of degeneracies. Namely, excitations which used to be in
the same irrep are now split into different irreps, which may have
different energies. This splitting of energy levels is
experimentally measurable, in principle.

To treat a phase with inversion symmetry, such as an achiral
tetrahedral nematic with $F(\Tdo)\times\C T_d$ symmetry (see
\cite{mbnematic:2006}), we need the formulae for electric symmetry
breaking of modified quantum doubles $F(H_m)\times\C H_{el}$. The
derivation of $\Rcal$ and $\Ucal$ after the condensation of a state
$|\phir>$ of an electric irrep of a modified quantum double
$F(H_m)\times\C H_{el}$ is analogous to the derivation given above
for a quantum double $D(H)$. The result is
\begin{eqnarray}
&&\Rcal = F(H_m)\tensor\C N_{\phir}\\
&&\Ucal = F(\Gamma^{-1}(N_{\phir}))\tensor\C N_{\phir}
\end{eqnarray}
For example, for $F(\Tdo)\tensor\C T_d$, $\Gamma^{-1}(N_{\phir}) =
\tilde{N_{\phir}}$, so that $\Ucal = F(\tilde{N_{\phir}})\tensor\C
N_{\phir}$.

As an example, we've worked out all possible electric breaking
patters from \mbox{$F(\Tdo)\tensor\C T_d$}, see table
\ref{tab:Tdsymmbreak}. Although we haven't found references that
systematically work out all electric condensates for all irreps as
we have done, the theory behind electric condensates is well known
and in that sense this is nothing new. Our
 interest was to rederive the results for ordinary condensates from the Hopf
symmetry description of liquid crystals. A reference that offers a
detailed analysis of the theory of ordinary condensates in bent-core nematic
liquid crystals is \cite{Lubensky:bent-core}, where a wealth of non-abelian phases
 are discussed. An analysis of the order
of the phase transitions using Renormalization group calculations is
done in \cite{Lubensky:rg}.

\begin{table}[h]
\begin{center}
\begin{tabular}{|c|c|c|c|}
\hline
                                &       &
                                & \\
Original symmetry               &  $T_d$ Irrep           & $\Rcal$   & $\Ucal$\\
                                & of condensate               &           &       \\
                                &       &
                                & \\
\hline\hline
                                &                                       &       &       \\
$F(\Tdo)\tensor\C T_d$  &$A_1$  &$F(\Tdo)\tensor\C T_d$ & $F(\Tdo)\tensor\C T_d$ \\
                    &$A_2$  &$F(\Tdo)\tensor\C T$ & $F(\Tdo)\tensor\C T$ \\
                    &$E$    &$F(\Tdo)\tensor\C D_2$&$F(\tilde{D_2})\tensor\C D_2$ \\
                    &$E$&$F(\Tdo)\tensor\C D_{2d}$&$F(\tilde{D_2})\tensor\C D_{2d}$\\
                    &$F_1$  &$F(\Tdo)\tensor\C S_4$&$F(\tilde{C_2})\tensor\C S_4$ \\
                    &$F_1$  &$F(\Tdo)\tensor\C C_{1v}$&$F(\tilde{C_1})\tensor\C C_{1v}$ \\
                    &$F_1$  &$F(\Tdo)\tensor\C C_3$&$F(\tilde{C_3})\tensor\C C_3$ \\
                    &$F_2$  &$F(\Tdo)\tensor\C C_{1v}$&$F(\tilde{C_1})\tensor\C C_{1v}$ \\
                    &$F_2$  &$F(\Tdo)\tensor\C C_{2v}$&$F(\tilde{C_2})\tensor\C C_{2v}$ \\
                    &$F_2$  &$F(\Tdo)\tensor\C C_{3v}$&$F(\tilde{C_3})\tensor\C C_{3v}$ \\
                                &       &                                       &
\\\hline
\end{tabular} \vspace*{.2cm}
\end{center}
\caption{{\small  Electric condensates in a tetrahedral nematic. We
use the notation for the irreps of $T_d$ given in \cite{Salt}, and
the standard crystallographic notation for groups (e.g. $S_4$ is not
the permutation group of 4 elements, it is a rotary-reflection
group).}} \label{tab:Tdsymmbreak}
\end{table}

As a side note, the transition from $F(\Tdo)\tensor\C T_d$ to
$F(\tilde{D_2})\tensor\C D_2$ in table \ref{tab:Tdsymmbreak},
induced by the condensate of a state in the irrep $E$ of $T_d$, is
an example of spontaneous symmetry breaking from an achiral to a
chiral phase, since $D_2$ does not contain any inversions or
reflections, while $T_d$ does. This may be the explanation of the
experimental discovery of a phase built up of achiral molecules,
whose symmetry is spontaneously broken to a chiral
phase\cite{Niori}. For a relevant discussion, see
\cite{Lubensky:bent-core}.

\subsection{Defect (magnetic) condensates}

There are different types of defect condensates which we wish to
analyze. Consider a phase described by a quantum double $D(H)$, or
a modified quantum double $F(H_m)\times\C H_{el}$, and pick a
magnetic representation $\Pi^A_{1}$ ($1$ is the trivial
representation of the centralizer $N_A$). A basis of the vector
space on which this irrep acts is given by $\{|g^A_i>\}$, where
the $g^A_i$ are the different defects in $A$. We consider the
following types of condensates:
\begin{itemize}
\item Single defect condensate
\begin{equation}
|\phi_r>=|g^A_i>
\end{equation}
\item Class sum defect condensate
\begin{equation}
|\phi_r>=\sum_{g_i \in A} |g_i> =: |C_{g_A}>
\end{equation}
where $C_{g_A}$ is a conjugacy class in the $D(H)$ case, and an
orbit in $G$ under the action of $H$ in the \mbox{$F(H_m)\times\C
H_{el}$} case. We denote the condensate by $|C_{g_A}>$, where
$g_A$ is the preferred element of $A$. \item Combined defect
condensate
\begin{equation}
|\phi_r>=\sum_{g_i\in E} |g_i>
\end{equation}
where $E$ is a subset of the defects in one class. We need only
take the elements to be within one class because, as we mentioned
earlier, we need only study the cases where the condensate is the
sum of vectors in the same irrep.
\end{itemize}

The single defect and class sum defect condensates are a special
case of combined defect condensate. The derivation of $\Rcal$ and
$\Ucal$ for a combined defect condensate is rather technical, so
we will discuss the results for the single defect and class sum
defect condensates first, and then derive the general formulae.

\subsubsection{Single defect condensate}

Consider a phase with $D(H)$ symmetry, and condense $|g_A>$ in the
magnetic irrep $\Pi^A_1$. We're condensing the chosen preferred
element in the conjugacy class $A$. This is not a restriction on
our choice of defect, since $g_A$ was chosen arbitrarily. The
condensate $|g_A>$ satisfies the trivial self braiding condition
(\ref{eq:TSB}).

The function $|\phi>: H\rightarrow\C$ that corresponds to the
vector $|g_A>$ is (see appendix \ref{app:Hopf})
\begin{equation}
|\phi(x)> = 1_{N_A}(x) \; \forall x\in H.
\end{equation}
The criterion (\ref{eq:Tresqd}) that defines $\Rcal$ becomes
\begin{eqnarray}
&& f \in \Rcal \nonumber \\
\iff \forall x,x_1,y_1 \in H: &&f(x_1 x g_A x^{-1},y_1) 1_{N_A}(y_1^{-1}x)=f(x_1,y_1)1_{N_A} (x) \nonumber \\
\forall x \notin N_A :&& f(x_1,y_1)=0 \quad \forall y_1 \notin N_A \nonumber \\
\forall x \in N_A :&& f(x_1 g_A,y_1) = f(x_1, y_1) \quad \forall y_1 \in N_A\nonumber \\
\Rightarrow && \Rcal = F(H/(g_A)) \otimes\C N_A
\label{eq:qdsingdef}
\end{eqnarray}
where we define $(g_A)$ to be the smallest subgroup of $H$ that
contains $g_A$.

This result for $\Rcal$ has a very natural interpretation: the
residual electric group is $N_A$, the subgroup of $H$ that doesn't
conjugate the defect. The magnetic part $H/(g_A)$ is not
necessarily a group. It consists of left cosets of $(g_A)=\{\ldots
g_A^{-1},e,g_A,g_A^2,\ldots\}$. The defects are now defined modulo
the condensate defect $|g_A>$. In other words, if a particle in a
magnetic irrep of the residual symmetry $\Rcal$ fuses with the
condensate $|g_A>$, it is left unchanged. Thus its defect is
defined modulo $g_A$.

Using our previous propositions, we can prove that $\Rcal$ is a
Hopf algebra $\iff$ $(g_A)$ is a normal subgroup of $H$ $\iff$
$H/(g_A)$ is a group.

The unconfined symmetry algebra is
\begin{equation}
\Ucal = D(N_A/(g_A)). \label{eq:Usinfdef}
\end{equation}
If we condense another defect $|kg_Ak^{-1}>$ in the conjugacy class
$A$, the symmetry algebras are$^{12}$ :
\begin{eqnarray}
&& \Rcal = F(H/k(g_A)k^{-1})\tensor\C kN_Ak^{-1} \\
&& \Ucal = D(N_A/k(g_A)k^{-1}).
\end{eqnarray}

The results for a single defect condensate $|g_A>$ in a phase with
$F(H_m)\times\C H_{el}$ symmetry are analogous:
\begin{eqnarray}
&& \Rcal = F(G/(g_A))\tensor\C N_A \label{eq:mqdTsingdef}        \\
&& \Ucal = F(N_A/(g_A))\tensor\C N_A/\Gamma((g_A)).
\end{eqnarray}

As an example, we've worked out all single defect condensates in
an achiral tetrahedral nematic in table \ref{tab:Tdsd}. For a
discussion of the derivation of these results, and for single
defect condensates in octahedral and icosahedral nematics, see
\cite{mbnematic:2006}.

\begin{table}[h]
\centering
\begin{tabular}{|c|c|c|c|}
\hline
                    &                   &                   &               \\
Single defect condensate in &       &               &           \\
$\Acal =F(\Tdo)\tens\C T_d $ & $K$   &$\Rcal$        &   $\Ucal$ \\
                    &                   &                   &               \\
\hline\hline
                &                   &                       &               \\
 $|-e>$         & $\tilde{C_1}$  &$F(T)\tensor\C T_d$    & $F(T)\tensor\C T_d$\\
 $|[123]>$      & $\tilde{C_3}$  &$F(T/C_3)\tensor\C C_3$& $D(e)$        \\
 $|[(12)(34)]>$ & $\tilde{C_2}$  &$F(T/C_2)\tensor\C D_2$& $D(C_2)$      \\
 $|-[(12)(34)]>$& $\tilde{C_2}$  &$F(T/C_2)\tensor\C D_2$& $D(C_2)$      \\
                    &                   &                   &               \\
\hline
\end{tabular} \vspace*{.2cm}
\caption{{\small  Single defect condensates in a tetrahedral
nematic.}} \label{tab:Tdsd}
\end{table}

\subsubsection{Class sum defect condensates}

Consider a phase with $D(H)$ symmetry, and condense the sum of the
defects in the conjugacy class $A$:
\begin{equation}
|\phi_r> = \sum_{g^A_i\in A} |g^A_i> =: |C_{g_A}>. \nonumber
\end{equation}
A class sum defect condensate satisfies the trivial self braiding
condition (\ref{eq:TSB}):
\begin{eqnarray}
\braid(|C_{g_A}>\tensor|C_{g_A}>)&=&\braid (\sum_{g^A_i\in A}
|g^A_i>\tensor \sum_{g^A_k\in A} |g^A_k>)
\nonumber\\
&=& \sum_{g^A_i\in A}(\sum_{g^A_k\in A} |g^A_i g^A_k (g^A_i)^{-1}>) \tensor |g^A_i> \nonumber\\
&=& \sum_{g^A_i\in A}(\sum_{g^A_k\in A} |g^A_k >) \tensor |g^A_i>  \nonumber\\
&=& |C_{g_A}>\tensor|C_{g_A}> . \nonumber
\end{eqnarray}
In going from the second to the third line, we use the fact that
$gAg^{-1}=A$ for any $g\in H$.

A class sum condensate doesn't break the electric group at all!
Namely, conjugation acts trivially on a conjugacy class, since for
any $g\in H$ we have
\begin{equation}
g\cdot |\phi_r> = g \cdot (\sum_{g^A_i\in A} |g^A_i>) =
\sum_{g^A_i\in A} |gg^A_ig^{-1}> = \sum_{g^A_i\in A} |g^A_i> =
|\phi_r>.
\end{equation}
Thus this condensate is invariant under all residual symmetry
transformations in $H$. For this reason, in the case of a local
gauge theory this condensate is  indeed the only physically
admissible \emph{gauge invariant magnetic condensate}. Namely, in
the $D(H)$ phase of a gauge theory, with $H$ a discrete group, the
only residual gauge transformations are global, because $H$ is
discrete and the gauge transformation must be continuously defined
on the space. These gauge transformations act trivially on the
class sum defect condensates, thus these condensates are indeed
gauge invariant.

The residual and unconfined symmetry algebras are
\begin{eqnarray}
&&\Rcal= F(H/K) \tensor \C H \\
&&\Ucal= D(H/K),
\end{eqnarray}
where $K$ is the smallest subgroup of $H$ that contains the class
$A$. From this definition, it follows that $K$ is a normal
subgroup of $H$. Thus $H/K$ is a group.

If we condense a class sum defect condensate $|C_{g_A}>$ in a
modified quantum double $F(H_m)\times\C H_{el}$, the outcome is
\begin{eqnarray}
&&\Rcal= F(H_m/K) \tensor \C H_{el} \\
&&\Ucal= F(H_m/K)\tensor\C H_{el}/\Gamma(K).
\end{eqnarray}

As an example, we've derived all class sum defect condensates in
an achiral tetrahedral nematic in table \ref{tab:Tdclsum}. The
class sum defect condensates in an achiral octahedral, and an
achiral icosahedral nematic are discussed in
\cite{mbnematic:2006}.

\begin{table}[t]
\centering
\begin{tabular}{|c|c|c|c|}
\hline
                    &                   &                   &               \\
Defect conjugacy classes of             &           &                   &               \\
$\Acal =F(\Tdo)\tens\C T_d $ & $K$       &$\Rcal$            &   $\Ucal$     \\
                    &                   &                   &               \\
\hline\hline
                    &                   &                   &               \\
 $|C_{-e}>$& $\tilde{C_1}$   &$F(T)\tensor\C T_d$& $F(T)\tensor\C T_d$ \\
 $|C_{[123]}>,|C_{-[123]}>,|C_{[124]}>,|C_{-[124]}>$    & $\Tdo$     &$\C T$ & $D(e)$\\
 $|C_{[(12)(34)]}>$& $\tilde{D_2}$   &$F(\Z_3)\tensor\C T_d$& $F(\Z_3) \tensor \C (\Z_3)_d$  \\
                    &                   &                   &               \\
\hline
\end{tabular} \vspace*{.2cm}
\caption{{\small  Class sum defect condensates in a tetrahedral
nematic. $(\Z_3)_d$ is isomorphic to the permutation group of 3
elements.}} \label{tab:Tdclsum}
\end{table}

\subsubsection{Combined defect condensates}

{\large The formal derivation}

We will now derive all the formulae for defect condensates we have
come across.

Start with a phase with $F(H_m) \tensor \C H_{el}$ symmetry.
Choose an irrep $\Pi^A_{\alpha}$, and consider a condensate of the
form $\sum_{g_i \in E} |g_i>$, with $E$ a subset of the defects in
one conjugacy class.

The demand of trivial self braiding (\ref{eq:TSB}) gives
\begin{eqnarray}
&&\braid(\sum_{g_i \in E} |g_i>\tensor\sum_{g_k \in E} |g_k>) =
\sum_{g_i \in E} |g_i> \tensor\sum_{g_k \in E}
|g_k>\nonumber\\
&\iff& \sum_{g_k \in E}\sum_{g_i \in E} |g_i g_k g_i^{-1}> \tensor
|g_i> = \sum_{g_i \in E} \sum_{g_k \in E}
|g_i> \tensor |g_k> \nonumber \\
&\iff& \forall g_i \in E : \{ g_i g_k g_i^{-1}\}_{g_k \in E} =
\{g_k\}_{g_k\in E}. \label{eq:TSBcomp}
\end{eqnarray}

It would be interesting in itself to construct the general solution
to this constraint, and to determine how many different defect
condensates satisfy this criterion. For example the
defect-antidefect condensate $|g>+|g^{-1}>$ always satisfies this
criterion$^{13}$, as does the superposition any set of commuting
elements in a certain conjugacy class, and class sum defect
condensates. This trivial self braiding condition will play a
crucial role in determining $\Rcal$.

The derivation of $\Rcal$ and $\Ucal$ is rather formal. We give
the results first, and then we derive them:
\begin{eqnarray}
&&\Rcal = F(H_m/K) \tensor \C M_E \nonumber \\
&&\Ucal = F(N_E/K)\tensor\C M_E/\Gamma(K). \nonumber
\end{eqnarray}

For the derivation, we must introduce various definitions. Define
the following subset of $H$ (which needn't be a subgroup):
\begin{equation}
V_E\subset H_{el}: \;V_E=\{x_i N_A\}_{g_i\in E}
\end{equation}
where $N_A \subset H_{el}$ is the normalizer of the chosen preferred
element $g_A$ in $A$, and $x_i$ satisfies $x_i g_A x_i^{-1}=g_i$. In
function notation, the condensate wave function is
\begin{equation}
|\phir(x)> = 1_{V_E}(x) \quad \forall x\in H.
\end{equation}
Define the following subgroup of $H_{el}$:
\begin{eqnarray}
M_E\subset H_{el}:\; M_E &=& \{m \in H: \{m\cdot g_i\}_{g_i\in E} = \{g_i\}_{g_i\in E} \}  \\
                     &=& \{m\in H_{el}: m V_E = V_E\}.
\label{eq:MEdef}
\end{eqnarray}
$M_E$ is composed of the global symmetry transformations that
leave the condensate invariant.

Also define
\begin{equation}
N_E\subset H_m: \; N_E = \{ n \in H_m: \{ng_in^{-1}\}_{g_i\in
E}=\{g_i\}_{g_i\in E}\}.
\end{equation}
Using (\ref{eq:Gammaone}): $\Gamma(g_1)\cdot g_2 = g_1 g_2
g_1^{-1}\;\forall g_1,g_2\in H_m$, we can prove that
\begin{equation}
\Gamma^{-1}(M_E) = N_E \quad \textrm{and}\quad \Gamma(N_E) = M_E.
\label{eq:MENE}
\end{equation}
From this equation we can derive that the elements of $M_E$ satisfy
\begin{equation}
|\phir(m x)> = |\phir(x)> \quad \forall m\in M_E, x \in H_{el}.
\label{eq:vedefs}
\end{equation}

Finally, we need one more definition:
\begin{equation}
K\subset H_m : \; K=(\{g_i\}_{g_i\in E}),
\end{equation}
where $(\{g_i\}_{g_i\in E})$ is the smallest subgroup of $H_m$
that all the $g_i\in E$, i.e. the defects in the condensate.

The trivial self braiding equation (\ref{eq:TSBcomp}) implies that
$K\subset N_E$. Thus, according to (\ref{eq:MENE})  and
(\ref{eq:vedefs})
\begin{equation}
\forall k \in K:\; \Gamma(k)\in M_E\quad \textrm{and}\quad
\phi(\Gamma(k) x) = \phi(x). \label{eq:Ktriv}
\end{equation}
The residual symmetry algebra $\Rcal$ is given by the set of
functions $f\in F(H_m\times H_{el})$ that satisfy
(\ref{eq:Tresqd}):
\begin{equation}
f(x_1 (x\cdot g_A),y_1) 1_{V_E} (y_1^{-1}x)=f(x_1, y_1)1_{V_E}(x).
\label{eq:Trescompeq}
\end{equation}
We will now prove that
\begin{equation}
\Rcal = F(H_m/K) \tensor \C M_E. \label{eq:Trescompres}
\end{equation}
Note that $H_m/K$ need not be a group.

To prove (\ref{eq:Trescompres}), take $y_1\notin M_E$. Then $
\exists\; x \in V_E $ such that $y_1^{-1}x\notin V$. Namely, if
such an $x$ doesn't exist, then $y_1^{-1}V=V$, thus $y_1^{-1}\in
M_E$ according to (\ref{eq:MEdef}), and $y_1\in M_E$.

Substitute an $y_1\notin M_E$, and $x$ with $y_1^{-1}x\notin V$,
into (\ref{eq:Trescompeq}). This gives
\begin{equation}
0 = f(x_1,y_1) \forall x_1\in H_m \quad \textrm{if}\quad y_1\notin
M_E,\nonumber
\end{equation}
so that $\Rcal \subset F(H_m)\tensor \C M_E$.

Now substitute $y_1\in M_E$ into (\ref{eq:Trescompeq}):
$1_{V_E}(y_1^{-1}x)=1_{V_E}(x)$ so the equation implies
$f(x_1(x\cdot g_A),y_1)=f(x_1,y_1)$ for all $x_1\in G, x\in V_E$.
Acting with all the $x\in V_E$ on $g_A$ gives us all the $g_i\in
E$, thus $f(x_1 g_i, y_1)=f(x_1,y_1)$ for all $g_i\in E$. Thus, in
the first component $f$ must be constant on left K cosets, since
$K$ is generated by the $g_i$. Thus $\Rcal = F(H_m/K)\tensor \C
M_E$.

$\Ucal$ is a little harder to extract. It is given by
\begin{equation}
\Ucal = F(N_E/K)\tensor\C M_E/\Gamma(K) \label{eq:Ucompdef}
\end{equation}
For the case of a quantum double $D(H)$
\begin{equation}
\Ucal = D(N_E/K). \label{eq:Ucalcompqd}
\end{equation}

To prove (\ref{eq:Ucompdef}) and (\ref{eq:Ucalcompqd}), we must
find out which irreps $\Omega^T_{\tau}$ of $F(H_m/K)\tensor\C M_E$
braid trivially with the condensate $\phi$ in the irrep
$\Pi^A_{\alpha}$ of $F(H_m)\times\C H_{el}$.

Our residual symmetry algebra $\Rcal$ is of the form
(\ref{eq:Rcalspec}), with $B=K$ and $N=M_E$. Thus we can use the
conditions (\ref{eq:Ucalone}) and (\ref{eq:Ucaltwo}) to determine
the irreps of $\Rcal$ that braid trivially with the condensate.
The unconfined symmetry algebra $\Ucal$ is then the Hopf algebra
whose irreps are precisely the unconfined irreps.

Equation (\ref{eq:Ucalone}) states that for an unconfined irrep
$\Omega^T_{\tau}$, with $g_T$ the preferred element in the orbit
$T$:
\begin{equation}
\sum_{k \in K} |\phir(\Gamma(k (x_{\eta}\cdot g_T^{-1}))x)> =
\sum_{k\in K} |\phir(\Gamma(k) x)> \quad \forall x \in H_{el},
\label{eq:Ucalfirst}
\end{equation}
where the $x_{\eta}$ are chosen representatives of left $N_T$
cosets in $M_E$.

Using (\ref{eq:Ktriv}), equation (\ref{eq:Ucalfirst}) becomes
\begin{eqnarray}
&&\sum_{k \in K} |\phir(\Gamma(x_{\eta}\cdot g_T^{-1})x)> =
\sum_{k\in K} |\phir(x)> \; \forall x \in H_{el}
\nonumber\\
&& \Rightarrow |\phir(\Gamma(x_{\eta}\cdot g_T^{-1})x)>=|\phir(x)> \;\forall x \in H _{el}\nonumber\\
&& \Rightarrow \Gamma(x_{\eta}\cdot g_T^{-1}) \in M_E \nonumber \\
&& \Rightarrow \Gamma(x_{\eta}\cdot g_T) \in M_E \nonumber \\
&& \Rightarrow x_{\eta}\cdot g_T \in N_E \quad \textrm{ using
(\ref{eq:MENE})}. \nonumber
\end{eqnarray}
Choosing $x_{\eta}= e$, we get $g_T\in N_E$. (\ref{eq:Ucalone}) is
actually equivalent to $g_T\in N_E$ in this case, because
$x_{\eta}\cdot g_T \in N_E$ follows from $g_T\in N_E$. To prove
this, note that $x_{\eta} \in M_E$, so $\{x_{\eta}\cdot
g_i\}_{g_i\in E}=\{g_i\}_{g_i\in E}$. Thus
\begin{eqnarray}
&&\{(x_{\eta}\cdot g_T) g_i (x_{\eta}\cdot g_T)^{-1}\}_{g_i\in E}
= \{(x_{\eta} \cdot g_T) (x_{\eta} \cdot g_i)
(x_{\eta} \cdot g_T)^{-1}\}_{g_i\in E} \nonumber \\
&& = \{(x_{\eta} \cdot (g_T g_i g_T)^{-1}\}_{g_i\in E} =
\{x_{\eta}\cdot g_i\}_{g_i\in E} = \{g_i\}_{g_i\in E}
\nonumber\\
&&\Rightarrow x_{\eta}\cdot g_T \in N_E  \nonumber
\end{eqnarray}
In proving the third equal sign we used the fact that $g_T \in
N_E$.

Thus (\ref{eq:Ucalone}) has taught us that for an irrep
$\Omega^T_{\tau}$ to be unconfined, we must have $g_T\in N_E$. The
magnetic part of the unconfined symmetry algebra $\Ucal$ is
therefore $F(N_E/K)$. From the definition of $N_E$ and $K$, we can
prove that
\begin{equation}
\forall n\in N_E : \; n K n^{-1} = K.
\end{equation}
Thus $K$ is a normal subgroup of $N_E$. $N_E/K$ is the unconfined
magnetic group.

Equation (\ref{eq:Ucaltwo}) further restricts $\Omega^T_{\tau}$:
\begin{eqnarray}
&&|\phir(x)>\neq 0, \;\Gamma(x\cdot g_A) \in M_E \nonumber\\
\Rightarrow &&\tau_{ij}(x_{\eta}^{-1} \Gamma(x\cdot g_A^{-1})
x_{\zeta}) 1_{N_T} (x_{\eta}^{-1} \Gamma( x\cdot g_A^{-1})
x_{\zeta}) = \tau_{ij}(x_{\eta}^{-1} x_{\zeta})
1_{N_T}(x_{\eta}^{-1} x_{\zeta}) \nonumber
\end{eqnarray}
Choose an $x$ such that $|\phir(x)>\neq 0$. This is equivalent to
saying that $x\cdot g_A = g_k$ for some $g_k\in E$. Now choose
$x_{\eta}=x_{\zeta}$ in (\ref{eq:Ucaltwo}):
\begin{equation}
\tau_{ij}(x_{\eta}^{-1} \Gamma(g_k^{-1}) x_{\eta}) 1_{N_T}
(x_{\eta}^{-1} \Gamma(g_k^{-1}) x_{\eta}) =
\tau_{ij}(e)=\delta_{ij}. \nonumber \label{eq:Ucaltwotau}
\end{equation}
Now $x_{\eta}^{-1} \Gamma(g_k^{-1}) x_{\eta} =
\Gamma(x_{\eta}^{-1}\cdot g_k^{-1}) = \Gamma((x_{\eta}^{-1}\cdot
g_k)^{-1})$. Since $x_{\eta} \in M_E$ and $g_k\in K$, we have
$(x_{\eta}^{-1} \cdot g_k) \in K$, so $(x_{\eta}^{-1}\cdot
g_k)^{-1} \in K$. Thus $\Gamma((x_{\eta}^{-1} g_k)^{-1})\in
\Gamma(K)$. Now $\Gamma(K)$ acts trivially on the magnetic group
$N_E/K$, due to (\ref{eq:Gammaone}). Thus necessarily
$\Gamma(K)\subset N_T$, since the elements of $\Gamma(K)$ are
normalizers of all elements of $N_E$, so they are also normalizers
of $g_T$. This means that $1_{N_T} (x_{\eta}^{-1}
\Gamma(g_k^{-1})x_{\eta})=1$.  \ref{eq:Ucaltwotau}) becomes
\begin{eqnarray}
&&\tau_{ij}(x_{\eta}^{-1} \Gamma(g_k^{-1}) x_{\eta}) =
\tau_{ij}(\Gamma(x_{\eta}^{-1} \cdot g_k^{-1})) =
\tau_{ij}(\Gamma(x_{\eta}^{-1} \cdot g_k))^{-1} = \delta_{ij}. \nonumber\\
&&\Rightarrow \tau_{ij}(\Gamma(x_{\eta}^{-1} \cdot g_k)) =
\delta_{ij}.\nonumber
\end{eqnarray}
Observe that the set $\{x_{\eta}^{-1} \cdot g_k\}_{\eta, k} = E$.
Since $\tau$ must send all $\Gamma(x_{\eta}^{-1} \cdot g_k)$ to
the unit matrix $\textbf{1}$, $\tau$ must send all of $\Gamma(K)$
to the unit matrix (since $K$ is generated by $E$). We conclude
that $\Gamma(K)$ is in the kernel of $\tau$, and the electric
group is $M_E/\Gamma(K)$.

At the start of this last derivation, we filled in
$x_{\eta}=x_{\zeta}$ in (\ref{eq:Ucaltwo}). The case $x_{\eta}
\neq x_{\zeta}$ gives nothing new, because
\begin{equation}
1_{N_T} (x_{\eta}^{-1} \Gamma(x\cdot g_A^{-1}) x_{\zeta})  =
1_{N_T} (x_{\eta}^{-1} \Gamma(x \cdot g_A^{-1})
x_{\eta}x_{\eta}^{-1}x_{\zeta} )=1_{N_T}(x_{\eta}^{-1} x_{\zeta}),
\end{equation}
where in the last line we used a fact that we proved earlier:
$x_{\eta}^{-1} \Gamma(x \cdot g_A)x_{\eta} \in N_T$. Thus
(\ref{eq:Ucaltwo}) becomes
\begin{eqnarray}
&& \tau_{ij} (x_{\eta}^{-1} \Gamma(x\cdot g_A^{-1}) x_{\zeta})
1_{N_T} (x_{\eta}^{-1} x_{\zeta}) = \tau_{ij}
(x_{\eta}^{-1} x_{\zeta}) 1_{N_T} (x_{\eta}^{-1} x_{\zeta}) \nonumber\\
&\iff& \tau_{ij} (x_{\eta}^{-1} \Gamma(x\cdot g_A^{-1})x_{\eta})
\tau_{ij}(x_{\eta}^{-1} x_{\zeta}) 1_{N_T} (x_{\eta}^{-1}
x_{\zeta}) = \tau_{ij}(x_{\eta}^{-1} x_{\zeta})
1_{N_T}(x_{\eta}^{-1} x_{\zeta}) \nonumber
\end{eqnarray}
This is equation is satisfied if $\tau_{ij} (x_{\eta}^{-1} \Gamma
(x\cdot g_A^{-1}) x_{\eta})$, which we already proved.

Summarizing, the unconfined magnetic group is $N_E/K$, and the
unconfined electric irreps are those that have $\Gamma(K)$ in
their kernel, which means that the electric group is $
M_E/\Gamma(K)$. Thus we have derived (\ref{eq:Ucompdef}). Had we
started with a quantum double $D(H)$ ($H=H_{el}=H_m$), the
unconfined symmetry algebra becomes $\Ucal = D(N_E/K)$, because in
that case $M_E = N_E$.

{\bf Acknowledgement:} F.A.B. would like to thank the Yukawa
Institute for Theoretical Physics in Kyoto, and in particular prof.
R. Sasaki for their hospitality. Part of the work was done while
visiting there.

\appendix
\section{Quasitriangular Hopf algebras}
\label{app:Hopf} In this appendix we summarize the mathematical
structure and notation connected to the Hopf algebras, which are
extensively used in this paper and in previous literature. The
applications focus on the quantum double $D(H)$ of a finite
(possibly non-abelian) group $H$. The essential information is
collected in a few comprehensive tables. In the text we merely
comment on the meaning and the physical interpretation/relevance
of the basic concepts. We conclude with a discussion of modified
quantum doubles.

Algebras are an ubiquitous mathematical structure. Another -
maybe less familiar  mathematical notion - is that of a
coalgebra, a structure that is in a precise sense a dual object
to an algebra. A Hopf algebra is simultaneously an algebra and a
coalgebra, with certain compatibility conditions between the two
structures. In the following subsections we systematically go
through some definitions and important examples.

\subsection{Algebras, coalgebras, and their duals}
An algebra $\mathcal{A}$  is a vector space over a field
$\mathbb{F}$ (which we will take to be $\mathbb{C}$), with a
bilinear multiplication.  We can think of the multiplication as a
map $ \mu: \mathcal{A}\tensor \mathcal{A} \mapsto \mathcal{A}$.
The algebras we discuss will all have a unit element, i.e. an
element $1$ which satisfies $1 a = a 1 = a \; \forall a \in
\mathcal{A}$. Thinking of it as a map one could say that the unit
embeds the field $\mathbb{F}$ into the center of the algebra. We
write $\eta: \mathbb{F} \mapsto \mathcal{A}$. We require that
$\eta$ be an algebra morphism (the field is also an algebra, with
itself as ground field). The algebras we consider are
associative with respect to their multiplication.

An important example for our purposes is the \emph{group algebra}
of a finite group. Label the groups elements by $h_i$.  The group
algebra is then the set of objects of the form $\sum_i \lambda_i
h_i$, with $\lambda_i\in \mathbb{C}$. This algebra is important,
as it contains all the information about the group. For example,
irreducible representations of the group algebra are in one-to-one
correspondence with irreducible  representations of the group.\\
In the physical models we consider, the regular (electric) modes
transform under an irrep of the group, which is equivalent to
saying that they transform under an irrep of the corresponding
group algebra.

A coalgebra $\Ccal$  is a vector space equipped with a
comultiplication and counit. The comultiplication is a linear map
$ \Delta: \Ccal \mapsto \Ccal \otimes \Ccal$.
The counit is a linear map $\epsilon: \Ccal \mapsto
\mathbb{F}$. They must satisfy the formal relations:
\begin{eqnarray}
\textrm{Coassociativity}: &&(\Delta\tensor id)\circ\Delta(c)=
(id\tensor\Delta)\circ \Delta(c) \forall c \in \Ccal\nonumber\\
\textrm{Counit}: && (\eps\tensor
id)\circ\Delta(c)=c=(id\tensor\eps)\circ\Delta(c) \forall c \in \Ccal\nonumber
\end{eqnarray}

We see in Table 4 that the group algebra naturally has a coalgebra
structure, where the comultiplication defines the action of the
group on the tensor product space.

Given an algebra $\Acal$, its dual$^{14}$ $\Acal^*$ can be given a
natural coalgebra structure. Similarly, given a finite dimensional
coalgebra, its dual has a algebra structure (in a slightly less
natural way reminiscent of the isomorphism $\Acal^{**}\simeq \Acal$
for a finite dimensional vector space). The structure of the duals
are given in Table 4, and the structures are given explicitly for
the group algebra and the function algebra, which are each
 other's duals. An example of a dual structure is the following: given a coalgebra
  $\Ccal$ with a coproduct $\Delta$, the dual $\Ccal^*$ has the following multiplication
  (which we denote by a star $\star$): $f_1 \star f_2 (x) \equiv (f_1 \star f_2) \circ \Delta (x)$.
  Thus the coproduct of $\Ccal$ is used to define the multiplication in $\Ccal^*$. Similarly,
  the counit of $\Ccal$ is used to define the unit in $\Ccal^*$.

\hoffset=-1.cm
\begin{table}[t]
\label{tab:ahopf}
 \fontsize{10pt}{12pt} \selectfont \centering
\begin{tabular}[h!]{|l|c|l||c|l|}
  \hline
  \multicolumn{3}{|c||}{}&
  \multicolumn{2}{|c|}{}\\
  \multicolumn{3}{|c||}{Hopf algebra $\mathcal{A}$}&
  \multicolumn{2}{|c|}{Dual Hopf algebra $\mathcal{A}^*$}\\
  \multicolumn{3}{|c||}{e.g.: Group algebra $\mathbb{C}H $}  & \multicolumn{2}{|c|} {Functions on the group $F(H)$ }\\
  \multicolumn{3}{|c||}{Basis: \{$h_i$\} \ $h_i \in H$}&
  \multicolumn{2}{|c|}{$\{f_i \}\;\; f_i = f_{h_i}= P_{h_i} \;\;\; f_i (x)= \delta_{h_i, x}  $}\\
  \multicolumn{3}{|c||}{}&
  \multicolumn{2}{|c|}{}\\
  \hline
  \hline
  & & & & \\
  & Algebra &$  $& Dual algebra &\\
  & & & & \\
  product &$ \cdot $&$ h_1 \cdot h_2=h_1 h_2 $&$ \star $&$ f_1 \star f_2 (x) = (f_1\otimes f_2) \circ \Delta(x) $ \\
  unit &$ e $&$ eh=he=h $&$ e^* $&$ e^*(x)= \varepsilon (x) = 1$ \\
  & & & & \\
  \hline
  & & & & \\
  & Co-algebra &$  $& Dual co-algebra &\\
  & & & & \\
  co-product &$ \Delta $&$ \Delta (h_i) = h_i\otimes h_i $&$ \Delta ^* $&$ \Delta^* (f)\; (x,y) = f(x \cdot y)$\\
  co-unit &$ \varepsilon $&$ \varepsilon (h) = 1 $&$ \varepsilon ^*$&$ \varepsilon^ * \;(f) = f (e)$ \\
  antipode &$ S $&$ S(h) = h^{-1} $&$ S^* $& $S^*(f)\; (x)= f(S(x)) = f(x^{-1}) $ \\
  & & & & \\
  \hline
  \end{tabular}
\vspace*{.2cm} \caption{{\small  The defining relations of the
group algebra and its dual, the algebra of functions on the group,
as Hopf algebras.}}
\end{table}
\hoffset=0.0cm

$F(H)$, the set of functions from $H$ to $\C$,  is the coalgebra
dual to the algebra $\C H$. It is also an algebra, and in fact it is
a Hopf algebra (see next section). These functions form a vector
space, which is spanned by the functions$^{15}$ $\{f_i = f_{h_i} :
h_i \in H\}$ with $f_i(x) = \delta_{h_i}(x)$. A coalgebra structure
is then defined by
\begin{eqnarray}
\label{eq:dualcoproduct1}
&&\eps(\delta_{h_i})=\delta_{h_i,e}\nonumber\\
&&\Delta(\delta_{h_i}) = \sum_{h_k \in H} \delta_{h_i h_k^{-1}}
\tensor \delta_{h_k}
\end{eqnarray}
We need only define these functions on a basis of $F(H)$ since they are linear.\\
Using the definition of the coproduct one arrives upon evaluation
on a pair of elements of $\Acal$ at,
\begin{eqnarray}
\label{eq:dualcoproduct2}
&&\Delta(f)(a_1,a_2)=f(a_1 a_2)\nonumber\\
&&\eps(f)=f(1).
\end{eqnarray}

\subsection{Hopf algebra's}
A Hopf algebra is simultaneously an algebra and a coalgebra, with
certain compatibility conditions. Namely, we demand that
$\epsilon$ and $\Delta$ be algebra morphisms. This is equivalent
to demanding that $\eta$ and $\mu$
be coalgebra morphisms (see \cite{Kassel}).\\
A Hopf algebra also contains an antipodal map $S$, which is the
unique map from $\Acal$ to $\Acal$ that satisfies
\begin{equation}
\mu(S\tensor id)\Delta(a)=\mu(id\tensor S)\Delta(a)
\end{equation}
From this definition, one can derive the following relations:
\begin{eqnarray}
\forall a, b \in \Acal:&& S(ab) = S(b)S(a) \\
&& S(e)=e \\
&&(S\tensor S) \Delta = \Delta^{op}S  \label{eq:antipode}\\
&& \eps(S(h))=\eps(h)
\end{eqnarray}
Physically, the antipode map is used to construct antiparticle
representations out of particle representations. This will become
clear if we are to talk about representation theory,  where given an
irrep $\Pi$, we define the \emph{antiparticle} or \emph{conjugate}
irrep as $\overline{\Pi}(a)=\Pi^t S(a)$, where $t$ denotes
transposition. $\overline{\Pi}$ is an irrep because $S$ is an
antimorphism, and so is the transposition. When we fuse particle and
antiparticle irreps, we don't necessarily get the vacuum
representation, but it is guaranteed that $\eps$ is present in the
fusion rules. This is completely analogous to the case of ordinary
groups.

The standard example of a Hopf algebra is the \emph{group
algebra}. We have already discussed the algebraic  and coalgebraic
structure. It is turned into a Hopf algebra by defining the
antipode $S$ as the inverse:
\begin{equation}
S(g_i)=g_i^{-1}
\end{equation}
So in fact all the familiar ingredients of a group(algebra) make
it already into a Hopf algebra.  Given a finite dimensional Hopf
algebra $\Acal$, one can quite generally define a Hopf algebra
structure on $\Acal^*$, and $\Acal^*$ is then called the dual Hopf
algebra. The definitions are quite natural and are given in Table
4. Note that for the dual Hopf algebra $F(H)=\mathbb{C}H^*$ the
antipodal map again involves the inverse, in the sense that
$S^*(f)(x)= f(x^{-1})$.

In the context of our physical applications, the dual version of the
algebra has very much to do with the topological defects and could
be called the ``magnetic'' part of the algebra. When we have
discussed the representation theory it will become clear what the
physical states are and what the action of the elements of the dual
algebra mean. One thing that may already be clear at this point is
that, if the topological charge corresponds to a group element, then
the comultiplication of the dual or function algebra determines the
action on the tensor product representation of the dual, and
therefore describes the fusion properties of the defects. It is then
clear from the expressions (\ref{eq:dualcoproduct1}) and
(\ref{eq:dualcoproduct2}) that the fusion indeed leads to the
required multiplication of group elements.

\subsection{Quasitriangularity}

A Hopf algebra $\Acal$ is called quasitriangular when there is an
invertible element $R$ of $\Acal\tensor\Acal$ that satisfies:
\begin{eqnarray}
(\Delta\tensor id)R = R_{13}R_{23} & (id\tensor\Delta)R=R_{13}R_{12} & \label{eq:r-element1}\\
\forall a \in \Acal: & R\Delta(a)R^{-1}=\Delta(a)&
\label{eq:r-element}
\end{eqnarray}
where if we write $R=\sum_{(R)}R^{(1)}\tensor R^{(2)}$, then
$$R_{ij}=\sum_{(R)}1\tensor\cdots\tensor R^{(1)}\tensor\cdots\tensor
R^{(2)}\tensor\cdots\tensor 1$$ where $R^{(1)}$ is in the i-th,
and $R^{(2)}$ in the j-th position.

The action of the R element on a tensorproduct of two
representations is to ``braid'' the two particles and is of crucial
importance to get the complete physical picture of particles and
defects which exhibit nontrivial braid properties. From the first
equations (\ref{eq:r-element1}) one derives that the Yang-Baxter
equation is satisfied. Equation (\ref{eq:r-element}) is the
important statement that the generators of the braid group commute
with the action of the Hopf algebra. This leads to the decomposition
of multi-particle states in product representations of the braid and
the Hopf algebra, thereby allowing for a clear definition of what we
mean by quantum statistics and nonabelian anyons, etc. We will
return to this subject shortly.

\subsection{The quantum double}
\hoffset=-1.0cm

\begin{table}[t!b]
\label{tab:quantumdouble} \fontsize{10pt}{12pt} \selectfont
\centering
\begin{tabular}{|l|c|l|}
  \hline
  \multicolumn{3}{|c|}{} \\
  \multicolumn{3}{|c|}{Double algebra  $\mathcal{D}= \mathcal{A}^* \tensor
  \mathcal{A}$} \\
  \multicolumn{3}{|c|}{Ex: Hopf double algebra $D(H)= F(H) \tensor \mathbb{C}H$}\\
  \multicolumn{3}{|c|}{ Basis: \{$f_i \times h_j$\} \ $f_i=P_{h_i} \in F(H), h_j \in H$}\\
  \multicolumn{3}{|c|}{} \\
  \hline
  \hline
   & & \\
  & Algebra &\\
  & &  \\
  product &$ \cdot $&$ (f_1 \times h_1) \cdot (f_2 \times h_2) (x)= f_1(x) f_2(h_1 x h_1^{-1}) \times h_1h_2 $  \\
  unit & $\mathbf{e}$& $(1 \times e) (x) = e $\\
  & &  \\
  \hline
  & &  \\
  & Co-algebra & \\
  & &  \\
  co-product &$ \Delta $&$ \Delta (f \times h)(x,y) = f(xy) h\otimes h $\\
  co-unit &$ \varepsilon $&$ \varepsilon (f \times h)(x) = f(e) $ \\
  antipode &$ S $&$ S (f \times h)(x) = f(h^{-1} x^{-1}h) h^{-1} $ \\
  & &  \\
  \hline
  & &  \\
  Central (ribbon) element& $ c $& $c= \sum_h (f_h \times h)$  \\
  R-element $R \in \mathcal{D}\otimes \mathcal{D}$ & $R$ & $R = \sum_h (f_h \times e) \otimes (1 \times h)$ \\
  & &  \\
  \hline
\end{tabular}
\vspace*{.2cm} \caption{{\small  The defining relations of the
quantum double $D(H)$ of a discrete group $H$.}}
\end{table}
\hoffset=0.0cm

Given a Hopf algebra $\Acal$, there is a natural way to ``double''
it, creating a new Hopf algebra $D(\Acal)$ called Drinfeld's quantum
double of $\Acal$. As a vector space, $D(\Acal)=\Acal^{*} \tensor
\Acal$, so it's a tensor product of $\Acal$ and its dual. For the
discussion of the Hopf algebra structure on $D(\Acal)$, see
\cite{Kassel}. For our purposes, we need only know what the
structure is like for $H$ a discrete group. We also
specify a braid matrix, making it a quasitriangular Hopf algebra.\\
As vector space, $D(H)$ is $F(H)\tensor \C H$, which is the same as
$F(H\times H)$ for finite $H$. Denote the basis elements of $D(H)$
as $(f_i\times h_j)$ or $P_h g$ (the latter is the notation used in
\cite{dwpb1995}), where $g\in H$, and $P_h\in F(H)$ is defined by
$P_h(x)=\delta_{h,x}$ ($x\in H$). \\If we look upon $P_h g$ as an
element of $F(H\times H)$, the function it corresponds to is $P_h g
(x,y) = \delta_{h,x}\delta_{g,y}$. We can also define $g \equiv 1 g
= \sum_{h\in H}P_h g$, and $P_h \equiv P_h e$. We've therefore
embedded $H$ and $F(H)$ into $D(H)$, and from these embeddings
obtained a basis of $D(H)$ (meaning that any element can be written
as a sum of products of these basis elements). The structure of
$D(H)$ is set by the following formulae:
\begin{eqnarray}
\label{eq:qdone}
&& P_h g P_{h'} g' = \delta_{h,gh'g^{-1}} P_h gg' \\
&& \Delta(P_h g) = \sum_{h'\in H} P_{hh'^{-1}} g \tensor P_{h'}g \\
&& \eps(P_h g) = \delta_{h,e} \\
&& S(P_h g) = P_{g^{-1}h^{-1} g} g^{-1} \\
\label{eq:qdfive}&& R = \sum_{g\in H} P_g e \tensor g
\end{eqnarray}

The complete structure is summarized in Table 5, including the
ribbon element $c$ (discussed later).

These relations can also be given in the language of functions in
$F(H\times H)$. The notation that comes along with this
formulation proves very convenient in actual calculations, so it
is useful to give the corresponding definitions here:
\begin{eqnarray}
\label{eq:qdfone}&& (f_1 f_2)(x,y) = \sum_{h\in H}f_1(x,h)f_2(h^{-1}xh,h^{-1}y) \\
&& \Delta(f)(x_1,y_1;x_2,y_2) = f(x_1 x_2,y_1)\delta_{y_1}(y_2)\\
&& \eps(f)=\sum_{z\in H}f(e,z)dz \\
&& S(f)(x,y) = f(y^{-1}x^{-1}y,y^{-1}) \\
&& R(x_1,y_1;x_2,y_2) = \delta_e(y_1)\delta_{x_1}(y_2)
\end{eqnarray}
This formulation is also important because it can be carried over
to define the quantum doubles  of continuous groups $H$, where $H$
is compact, or locally compact. For results on such cases, see
\cite{Koornwinder:loccpt} and \cite{Koornwinder:cpt}.

We finally should mention that the quantum double has a central
element, denoted by $c$ and called the ribbon element. Its
eigenvalue can be used as a label on the representations of the
$D(H)$. In applications it defines the phase obtained after
rotating the particle state by $2\pi$ and is therefore called the
\emph{spin factor} of a given representation. It plays a central
role in defining a suitable generalization of the spin statistics
connection, to which we turn next.
\subsection{A non-Abelian spin statistics connection}

\begin{figure}[t]\label{fig:spinstatcon}
\begin{center}
\scalebox{0.9}{\includegraphics*[0.8in,1in][8in,4.9in]{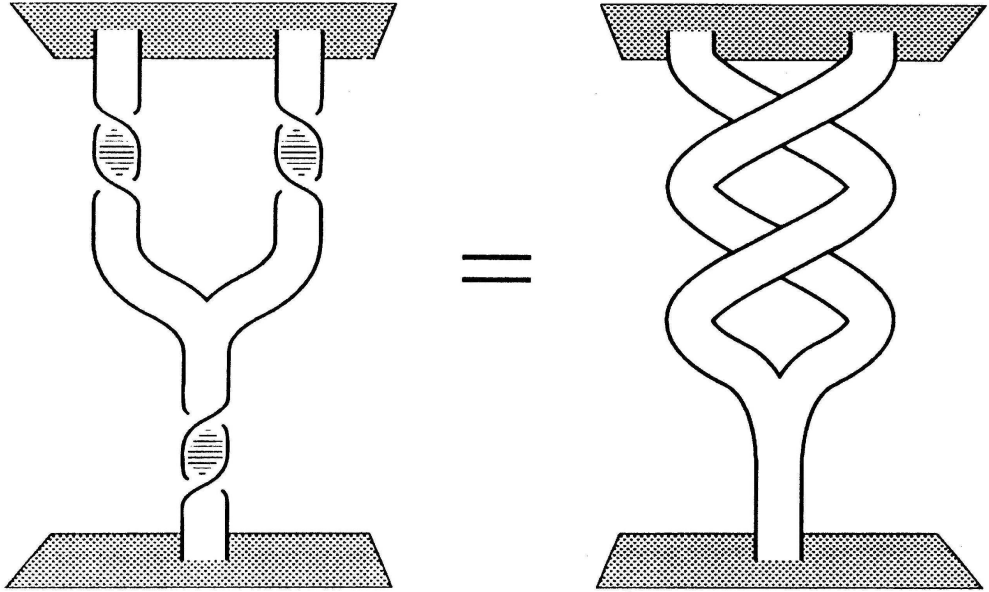}}
\caption{{\small  The suspender diagram gives the spin statistics
relation in the case of nontrivial braid statistics. The equal sign
denotes the topological equivalence between both pictures: unwinding
the right hand side yields the left hand side. There is a direct
relation with equation (\ref{spinstatcon}).  The rotation of the
"ribbons" is the result of acting with the central ribbon element
$c$, while the braiding of the ribbons represents the action of $R$.
The splitting represents the action of $\Delta$ .}}
\end{center}
\end{figure}

Let us briefly return to the important role played by the $R$
element which as mentioned, generates the action of the braid group
on a multi-particle states corresponding to some state in a tensor
product of representations of $D(H)$. We noted already that the
braiding with $R$ commutes with action of $D(H)$:
\begin{equation}
  \Delta^{op} (f \times h)R = R \Delta (f \times h)
\end{equation}
where $\Delta^{op} \equiv \tau \Delta$ (i.e. $\Delta$ followed by a
trivial permutation $\tau$ of the two strands). From this it follows
that the $n$-particle states form representations of $D(H) \otimes
B_n$, with $B_n$ the braid group on $n$ strands. We speak of
non-abelian statistics if the theory realizes states that correspond
to higher dimensional representations of $B_n$. In this context it
is an important question to what extent one can still speak of a
spin-statistics connection. One can indeed write down a generalized
spin-statistics theorem  in terms of the action of the braid group
end the central ribbon element which corresponds to physically
rotating the defect over an angle $2\pi$ and therefore generating
the phase factor due to the (fractional) spin of the particle in
question. It reads:
  \begin{equation}\label{spinstatcon}
  (c\otimes c) \Delta(a) c^{-1} = (R_{21} R_{12}) \Delta(a) \;
  \forall a \in \Acal,
  \end{equation}
and can be represented graphically by the ``suspenders diagram''
depicted in Figure 6. Note that the concept of a spin statistics
connection has become considerably more intricate. It is no longer
an attribute carried by a single particle. It may involve two
different particles (mutual statistics) and is also dependent on the
channel in which they are fused.

\subsection{Representation theory}
The irreducible representations of $D(H)$ follow from the
observation that it's a transformation group algebra (see
\cite{Bais:2002ny} for details). Since we will come across other
transformation group algebras in the main text, we give the general
definition. As a vector space, the transformation group algebra is
$F(X)\tensor\C H$, where $X$ is a finite set and $H$ is a finite
group. A basis is given by
\begin{equation}
\{P_x h : x\in X, h\in H\}.
\end{equation}
Just as in the quantum double case, we can consider it to be the
vector space $F(X\times H)$ of functions
\begin{equation}
f: X\times H \rightarrow \C.
\end{equation}
The element $P_xh\in F(X)\tensor\C H$ corresponds to the function
\begin{equation}
f(y,z)= \delta_{x,y}\delta_{h,z} \; \forall y\in X, z\in H.
\end{equation}
Furthermore, there is an action of the group $H$ on $X$. This
means that the elements $h\in H$ act as bijections of $X$, in a
manner that is consistent with the group structure (i.e. we have a
homomorphism from $H$ to bijections of $X$). Denote the action of
$h\in H$ on $x\in X$ by $h\cdot x$. We now turn this vector into
an algebra, by introducing a multiplication.
\begin{defn}
$F(X\times H)$ is a transformation group algebra if the
multiplication of $P_x h$ and $P_{x'}h'$ is given by:
\begin{equation}
P_x h P_{x'}h' = \delta_{x,h\cdot x'} P_x hh'
\end{equation}
In function notation:
\begin{equation}
(f_1 f_2)(x,y) = \sum_{h\in H}f_1(x,h)f_2(h^{-1}\cdot x,h^{-1}y)
\end{equation}
\end{defn}

We define an \emph{inner product} on $F(X\times H)$. We give it in
both notations: in terms of elements $P_xh\in F(X)\tensor\C H$,
and in terms of functions $f_i\in F(X\times H)$.
\begin{eqnarray}
&&(P_x h, P_{x'}h') = \delta_{x,x'}\delta_{h,h'}\nonumber\\
&&(f_1, f_2) = \sum_{x\in X, h\in H}f_1(x,h)^*f_2(x,h).
\label{eq:inner}
\end{eqnarray}

We can split $X$ up into orbits under the action of $H$. The orbit
of an $x_0 \in X$ is given by $\{h\cdot x_0: h\in H\}$. Call
$\{A\}$ the collection of orbits. For each orbit $A$, choose a
preferred element $x^A_1$, and define the \emph{normalizer} $N_A$
to be the subgroup of $h\in H$ that satisfy $h\cdot x^A_1 =
x^A_1$. The $O_A$ and $N_A$ play a central role in the
determination of the irreps.
\begin{thm}
Choose an orbit $A$ in $X$, a preferred element $x^A_1$ of $A$,
and an irrep $\alpha$ of $N_A$. The orbit
$A=\{x^A_1,x^A_2,\ldots,x^A_n : x^A_i\in X\}$. Let $h^A_i$ be any
element in $H$ such that $h^A_i\cdot x^A_1 = x^A_i$. Then the
$h^A_i$ form representatives of left $N_A$ cosets in $H$. Further
call $e_j$ the basis vectors
of the vector space $V_{\alpha}$ on which the irrep $\alpha$ acts. \\
An irreducible unitary representation $\Pi^A_{\alpha}$ of
$F(X\times H)$ is given by inducing the irrep $\alpha$. A basis of
the vector space is $\{|x^A_i,e^{\alpha}_j>\}$, and the action of
$P_x g \in F(X\times H)$ is given by
$$\Pi^A_{\alpha}(P_x g) |x^A_i,e^{\alpha}_j> =
\delta_{x,gh^A_i\cdot x^A_1} |x^A_j,\alpha(n)e^{\alpha}_j>  $$
where $x^A_j$ is defined by $gh^A_i=h^A_j n, n \in N_A$. This is
possible because $gh^A_i$ sits in a particular
coset of $H/N_A$, and the $h^A_i$ are representatives of the left $N_A$ cosets.\\
Furthermore, all unitary irreducible representations are
equivalent to some $\Pi^A_{\alpha}$, and $\Pi^A_{\alpha}$ is
equivalent to $\Pi^B_{\beta}$ iff $O_A=O_B$ and $\alpha$ is
equivalent to $\beta$.
\end{thm}
Thus irreps of a transformation group algebra are labelled by an
orbit $A$ in $X$ under the action of $H$, and an irrep $\alpha$ of
the normalizer $N_A$ of a chosen preferred element $x^A_1$ of $A$.

The notation in terms of basis elements makes the action of the
irreps transparent. An alternate notation for the Hilbert space is
$h^A_i \tensor |e^{\alpha}_j> \equiv |x^A_i,e^{\alpha}_j>$. Then
the action of a global symmetry transformation $g\equiv 1g$ is
simply
$$\Pi^A_{\alpha}(g)h^A_i \tensor |e^{\alpha}_j>=gh^A_i \tensor |e^{\alpha}_j>=h^A_j n \tensor
|e^{\alpha}_j>=h^A_j \tensor |\alpha(n)e^{\alpha}_j>\equiv
|x^A_i,\alpha(n)e^{\alpha}_j>$$ In words, the part of g that
"shoots through" the defect acts on the electric part.

\hoffset=-1.0cm
\begin{table}[tb]
\fontsize{10pt}{12pt} \selectfont \centering
\begin{tabular}{|l|c|l|}
  \hline
  \multicolumn{3}{|c|}{} \\
  \multicolumn{3}{|c|}{Representations $\Pi^A_\alpha$ of $D(H)= F(H) \times \mathbb{C}H$} \\
  \multicolumn{3}{|c|}{} \\
  \hline
  \hline
   & & \\
  representation  &$ \Pi^A_\alpha $& $ A \sim$ defect/magnetic label, $\alpha \sim$ ordinary/electric label   \\
  & $A$ &  $C_A \sim $ Conjugacy class (orbit of representative element
  $h_A$).\\
  &$\alpha$ & $\alpha \sim$ is a representation of the normalizer
  $N_A$ of $h_A$ in $H$.\\
  & & \\
  carrier space & $V^A_\alpha$ & $ |v>: H\rightarrow V_\alpha \;\;\{|v(x)>\; | \; |v(xn)>=\alpha(n^{-1})\; |v(x)>,\; n \in N_A \} $\\
  & &  \\
  action of $D(H)$ on $V^A_\alpha$ &   & $\pi^A_\alpha (f \times h) |v(x)> = f(xhx^{-1})\;|v(h^{-1}x)>$  \\
  &  &  \\
  \hline
  & &  \\
  central element & $c$ &  $\Pi^A_\alpha (c) |v(x)> = \alpha(h_A^{-1}) |v(x)>$\\
  spin factor & $s^A_\alpha $ & $s^A_\alpha \equiv \alpha (h_A^{-1})$ \\
  & &  \\
  \hline
  &  &  \\
  tensor products &$ \Pi^A_\alpha \otimes \Pi^B_\beta $ &
  $\Pi^A_\alpha \otimes \Pi^B_\beta (f\times h) V\otimes W \equiv
  \Pi^A_\alpha \otimes \Pi^B_\beta \Delta (f\times h) V\otimes W$ \\
   &  &  \\
  &  & Clebsch-Gordan series:  \\
   &  & $ \Pi^A_\alpha \otimes \Pi^B_\beta = \sum_{C, \gamma} N_{\alpha \beta C}^{A B \gamma}\; \Pi_\gamma^C $ \\
   &  &  \\
  \hline
\end{tabular}
\vspace*{.2cm} \caption{{\small  The defining properties for the
representations $\Pi^A_\alpha$ of the quantum double $D(H)$ of a
discrete group $H$.}}
\end{table}
\hoffset=0.0cm

The function notation is rather opaque, but extremely useful in
calculations in the main text. The Hilbert space of the irrep
$\Pi^A_{\alpha}$ is given by:
$$F_{\alpha}(H,V_{\alpha})=\{|\phi>:H\rightarrow V_{\alpha}||\phi(hn)>=
\alpha(n^{-1})|\phi(h)>,\forall h\in H, \forall n\in N\}$$ To make
contact with the notation above, $|\phi(h)>$ corresponds to the
vector attached to the ``flux'' $h\cdot x^A_1$. For example, the
function $|\phi>$ associated with $|x^A_i,e^{\alpha}_j>$ is
$|\phi(h^A_i)> = 1_{h^A_i N_A}(h) |e^{\alpha}_j>$. To explain the
rest of the definition, note that $\forall n \in N$
$$|x^A_i,e^{\alpha}_j> \equiv h^A_j\tensor|e^{\alpha}_j>=
h^A_jnn^{-1}\tensor|e^{\alpha}_j>
=h^A_jn\tensor|\alpha(n^{-1})e^{\alpha}_j>  $$
which explains why $|\phi(xn)>=\alpha(n^{-1})|\phi(x)>$.\\
Then the action of $f\in F(X\times H)$ on $|\phi>$ under the irrep
$\Pi^A_{\alpha}$ gives a new function $\Pi^A_{\alpha}(f)\phi\in
F_{\alpha}(H,V_{\alpha})$, defined by
$$(\Pi^A_{\alpha}(f)\phi)(h)=\sum_{z\in H} f(h\cdot x^A_1,z)|\phi(z^{-1}h)>$$
One easily checks that this is equivalent to the definition given
above.

The quantum double $D(H)$ is a special case of a transformation
group algebra, with $X=H$ and $h\cdot x = hxh^{-1}$. Thus in
$D(H)$ the orbits of $X=H$ under the action of $H$ are conjugacy
class of $X=H$. $N_A$ is the centralizer of the preferred element
$g_A$ of $A$.

We proved earlier on that the antipode can be used to create an
antiparticle irrep from any irrep $\Pi^A_{\alpha}$. In the case of
$D(H)$, the antiparticle irrep of $\Pi^A_{\alpha}$ is
$\Pi^{A^{-1}}_{\overline{\alpha}}$, where $A^{-1}$ is the
conjugacy class of $g_A^{-1}$, and $\overline{\alpha}(n) =
\alpha^t (n^{-1})$. In particle, for an electric irrep
$\Pi^e_{\alpha}$ the antiparticle irrep is
$\Pi^e_{\overline{\alpha}}$. An example of this is the \textbf{3}
irrep of $SU(3)$, which is the antiparticle irrep of \textbf{3}.
The quarks transform under the \textbf{3} irrep, while the
antiquarks transform under the \textbf{$\overline{3}$} irrep of
$SU(3)$.

\subsection{Modified quantum doubles}

In the main text we came across phases whose Hopf symmetry was a
variation on a quantum double, which we called a \emph{modified
quantum double}. We will now give the general definition of a
modified quantum double$^{16}$.

As a vector space a modified quantum double is $F(H_m)\tensor\C
H_{el}$, where $H_m$ and $H_{el}$ are two groups. It is also a
transformation group algebra, so there is an action of $H_{el}$ on
$H_m$. We denote the action of $h\in H_{el}$ on $g\in H_m$ by
$h\cdot g$. We require this action to satisfy the following
relation:
\begin{equation}
\forall h\in H_{el},\; \forall g,g' \in H_m: h\cdot(gg') = (h\cdot
g)(h\cdot g'). \label{eq:modqdact}
\end{equation}
A basis of $F(H_m)\times\C H_{el}$ is given by $\{P_gh:g\in
H_m,h\in H_{el}\}$. $F(H_m)\times\C H_{el}$ is a Hopf algebra
with the following structure:
\begin{eqnarray}
\forall g,g'\in H_m, \forall h,h'\in H_{el}:&& P_ghP_{g'}h'=\delta_{g,h\cdot g'} P_ghh'\\
                                    && \Delta(P_gh)=\sum_{g'\in H_m}P_{g'}h \tensor P_{g'^{-1}g}h \\
                                    && \eps(P_gh) = \delta_{g,e} \\
                                    && S(P_gh) = P_{h^{-1}\cdot g^{-1}}h^{-1}
\end{eqnarray}
These structures satisfy all the axioms of a Hopf algebra, thus
$F(H_m)\times\C H_{el}$ is a Hopf algebra. Just as in the quantum
double case, we can write elements of $F(H_m)\times\C H_{el}$ as
functions in $F(H_m \times H_{el})$. An element of $F(H_m \times
H_{el})$ is a function
\begin{equation}
f: H_m\times H_{el} \rightarrow \C.
\end{equation}

The element $P_gh \in F(H_m)\times\C H_{el}$ corresponds to the
function
\begin{equation}
f(y,z) = \delta_{g,y}\delta_{h,z} \; \forall y\in H_m, z\in
H_{el}.
\end{equation}
The structures of $F(H_m\times H_{el})$ are
\begin{eqnarray}
&&\forall f, f_1, f_2 \in F(H_m \times H_{el}); x, x_1, x_2\in H_m; y, y_1, y_2 \in H_{el}: \nonumber\\
&& (f_1\times f_2)(x,y)=\sum_{h\in H_{el}} f_1(x,h)f_2(h^{-1}\cdot x, h^{-1}y) \\
&& \Delta(f_1)(x_1,y_1;x_2,y_2)=f(x_1x_2,y_1)\delta_{y_1}(y_2)\\
&&\eps(f) = \sum_{h\in H_{el}}f(e,h)dh\\
&&S(f)(x,y)=f(y^{-1}\cdot x^{-1},y^{-1})
\end{eqnarray}

Making $F(H_m)\times\C H_{el}$ quasitriangular, i.e. introducing
a braid matrix, is not that easy. We have found a way to do it
that applies to all physical systems we looked at. We need a
homomorphism:
\begin{equation}
\Gamma: H_m \rightarrow H_{el},
\end{equation}
that satisfies the following relations:
\begin{eqnarray}
&&\forall g, g_1, g_2 \in H_m, h \in H_{el}:\\
&& \Gamma(g_1)\cdot g_2 = g_1 g_2 g_1^{-1} \label{eq:Gammaone}\\
&& \Gamma(h\cdot g) = h\Gamma(g)h^{-1}. \label{eq:Gammatwo}
\end{eqnarray}
$F(H_m)\times\C H_{el}$ is then a quasitriangular Hopf algebra
with the following braid matrix:
\begin{equation}
R=\sum_{g\in H_m}P_g e \tensor \Gamma(g).
\end{equation}
The inverse of $R$ is
\begin{equation}
R^{-1}=\sum_{g\in G}P_g e \tensor \Gamma(g^{-1}).
\end{equation}

The quantum double $D(H)$ is a special case of a modified quantum
double, with $H_m=H_{el}=H$, $h\cdot g = hgh^{-1}\;\forall h,g\in
H$, and $\Gamma\equiv id$.
\section{ Defining a coproduct for $\Rcal$}
$\Rcal$ is not always a Hopf algebra, because the coproduct of
$\Acal$ does not necessarily satisfy $\Delta(\Rcal)\subset
\Rcal\tensor\Rcal$. It does satisfy $\Delta(\Rcal) \subset \Acal
\tensor \Rcal$. This means that we cannot take the tensor product
of irreps of $\Rcal$, in other words we cannot fuse two particles
in $\Rcal$. We can only fuse particles in $\Rcal$ with particles
of $\Acal$ coming in from the left.

The reason we can't fuse particles of $\Rcal$ is that some
excitations of the condensate $|\phir>$ are confined. The
condensate takes on the value $|\phir>$ to the right, and
$|\phil>\neq |\phir>$ to the left of a confined excitation, in an
irrep $\Omega$ of $\Rcal$. Thus particles coming in from the left
see the condensate $|\phil>$. Thus particles to the left of are
excitations of $|\phil>$, and they transform under irreducible
representations of the right residual symmetry algebra of
$|\phil>$.

Matters are complicated further by the observation that given two
states $|v_1>$ and $|v_2>$ in a confined irrep $\Omega$, the value
of the condensate to the left of $|v_1>$ need not be equal to the
value the condensate takes to the left of $|v_2>$. Thus the
condensate does not take a well defined value to the left of the
irrep $\Omega$.

One possible interpretation of a configuration with a confined
excitation $\Omega$, is that the condensate has only condensed to
the right of $\Omega$. To the left of $\Omega$ the system is in
the unbroken phase, with $\Acal$ symmetry. Thus particles to the
left of $\Omega$ are irreps of $\Acal$. We can fuse particles in
irreps $\Pi$ of $\Acal$ with $\Omega$, and look at the fusion
rules. The outcome of this fusion tells us what the particle in
the irrep $\Acal$ can become when it enters the ordered phase.

This interpretation of a confined excitation is reasonable, but it
is unsatisfactory. Namely, we still want to be a able to fuse the
irreps of the residual symmetry algebra. There should be a finite
set of quantum numbers in the broken phase, and we should be able
to tell which of these quantum numbers can form hadronic
composites. We want to be able to talk about the fusion of the
quantum numbers in the broken phase without bringing in the
quantum numbers of $\Acal$. We know that if $\Rcal$ is a Hopf
algebra, we can fuse irreps of $\Rcal$. We will now discuss how to
fuse irreps of $\Rcal$ when $\Rcal$ is not a Hopf algebra.

\subsection{A purely magnetic phase}

Let us first study a simple system that only has magnetic degrees
of freedom. Its symmetry algebra is $F(H)$, with $H$ some finite
group. A basis of $H$ is given by $\{P_h : h\in H\}$, where $P_h$
is a projection operator. It measures the flux of a configuration
: if the flux of a state $|v>$ is $h$, then $P_h\cdot|v>=|v>$. If
the flux of $|v>$ isn't $h$, then $P_h|v>=0$.

Now consider a single defect condensate $|\phir>=|g>$. Then using
(\ref{eq:mqdTsingdef}) (with $N_A=\{e\}$) we find for the residual
symmetry algebra:
\begin{equation}
\Rcal = F(H/(g)),
\end{equation}
where $(g)$ is the subgroup of $H$ generated by $|g>$. Let us
assume that $(g)$ is not a normal subgroup of $H$, so that $\Rcal$
is not a Hopf algebra. A basis of $\Rcal$ is given by $\{
P_{h_i(g)} : h_i \in F \}$, where $F$ consists of a set $\{h_i\}$
of representatives of the left $(g)$ cosets in $H$. These
projection operators measure defects modulo the condensed defect
$|g>$. This is a consequence of our definition of operators in
$\Rcal$: the operators in $\Rcal$ are the operators that do not
notice when a particle fuses with the condensate $|g>$.

If we apply the coproduct $\Delta$ on a basis element $P_{h_k(g)}$
of $\Rcal$, we obtain
\begin{equation}
\Delta(P_{h_k(g)}) = \sum_{h_i \in F} P_{h_k (g) h_i^{-1}} \tensor
P_{h_i(g)} = \sum_{h_i \in F} P_{h_kh_i^{-1} (h_i g h_i^{-1})}
\tensor P_{h_i(g)}.
\end{equation}
This coproduct has a natural interpretation: if a projection
operator $P_{h_i(g)}$ measures a defect on the right, then the
projection operators on the left measure defects modulo
$h_i(g)h_i^{-1}$. Thus $\phil=|h_igh_i^{-1}>$: the condensate on
the left is conjugated by $h_i$. Note that the defect of the
particle that $P_{h_i(g)}$ measured is only defined modulo g, but
this doesn't affect $|\phil>$, since $| h_ig^n g
(h_ig^n)^{-1}>=|h_igh_i^{-1}> \; \forall n\in\Z$.

We assume that there is a confined excitation $|v>$ in an $\Omega$
of $\Rcal$, and that $P_{h_i(g)} |v>=|v>$. We have a condensate
$|\phil>$ to the left of the confined excitation measured by
$P_{h_i(g)}$ that is different from the condensate $|\phir>$ to
the right. We can, however, \emph{redefine our projection
operators to the left of the confined excitation}. The projection
operators to the left of $|v>$ are defined modulo
$|h_igh_i^{-1}>$. A basis of these projection operators is $\{
P_{h'_k(h_igh_i^{-1})} : h'_k \in J \}$, where $J$ consists of a
set $\{h'_k\}$ of representatives of the left $(h_igh_i^{-1})$
cosets in $H$.

To every projection operator $P_{h'_k(h_i g h_i^{-1})}$, we
associate a projection operator $P'_{h_i(g),h'_k (g)}$ in $\Rcal$:
\begin{equation}
P'_{h_i(g),h'_k (g)} = P_{ h_i^{-1}  h'_k(h_i g h_i^{-1}) h_i} =
P_{ h_i^{-1}h'_kh_i (g)}.
\end{equation}

Using this definition, we can define a coproduct $\Delta'$ for
$\Rcal$:
\begin{equation}
\Delta'(P_{h_k(g)}) = \sum_{h_i\in F} P'_{h_i(g),h'_k (g)} \tensor
P_{h_i (g)} = \sum_{h_i\in F} P_{ h_i^{-1}h'_kh_i (g)} \tensor
P_{h_i(g)}.
\end{equation}

$\Delta'$ is a map from $\Rcal$ to $\Rcal\tensor\Rcal$. One can
check that $\Delta'$ is an algebra morphism, i.e.
\begin{equation}
\Delta'( P_{h_j(g)} P_{h_k(g)}) = \Delta'( P_{h_j(g)}) \Delta'(
P_{h_k(g)}).
\end{equation}

We can use $\Delta'$ to fuse two irreps $\Omega_2$ and $\Omega_1$
of $\Rcal$:
\begin{equation}
\Omega_2\tensor\Omega_1 (a) =
(\Omega_2\tensor\Omega_1)\circ\Delta'(a).
\end{equation}
$\Delta'$ is not coassociative, i.e.
\begin{equation}
(\Delta'\tensor id)\circ\Delta' \neq
(id\tensor\Delta')\circ\Delta'.
\end{equation}
This implies that the tensor product of three irreps $\Omega_3$,
$\Omega_2$ and $\Omega_1$ of $\Rcal$ is not associative:
\begin{equation}
(\Omega_3 \tensor \Omega_2) \tensor \Omega_1 \neq \Omega_3 \tensor
(\Omega_2 \tensor \Omega_1).
\end{equation}
The interpretation of this non coassociativity is as follows: when
we take the tensor product $\Omega_2 \tensor \Omega_1$, $\Omega_2$
is defined \emph{with respect to the condensate to the right of
$\Omega_1$}. If a third particle $\Omega_3$ comes in from the
left, then it becomes defined with respect to the condensate
\emph{between $\Omega_2$ and $\Omega_1$}. So we have a natural
ordering for the tensor product, namely we must take $(\Omega_3
\tensor \Omega_2) \tensor \Omega_1$, which corresponds to having
$\Omega_1$ in the system, then bringing in $\Omega_2$ from the
left, and then bringing in $\Omega_3$. $\Omega_3 \tensor (\Omega_2
\tensor \Omega_1)$ is unphysical, because it isn't clear how the
particles were put in the system.

\subsection{The general case}

In general, if we bring in $\Omega_1, \Omega_2, \ldots , \Omega_n$
from the left in that order, the resulting configuration is
$((\ldots (\Omega_n \tensor \Omega_{n-1}) \tensor \Omega_{n-2})
\tensor \ldots )\tensor \Omega_2 )\tensor \Omega_1$. We have been
forced to introduce an ordering in our fusion. This ordering
corresponds to an ordering in the coproduct:
\begin{equation}
(\Delta'\tensor id \tensor \ldots \tensor id) \circ \ldots \circ
(\Delta' \tensor id \tensor id) \circ (\Delta' \tensor id)\circ
\Delta'(a).
\end{equation}

Once we have defined a non coassociative coproduct $\Delta'$, we
can fuse confined excitations $\Rcal$, and study the possible
hadronic composites. To define a $\Delta'$, we need a linear map
$\gamma$ with the following properties:
\begin{eqnarray}
&&\gamma : \Acal \tensor \Rcal \mapsto \Acal \tensor \Rcal \nonumber \\
&&\forall a,b, \in \Rcal: \gamma(\Delta(ab)) = \gamma(\Delta(a))\gamma(\Delta(b)) \nonumber \\
&&\gamma |_{\Delta(\Rcal)} : \Delta(\Rcal) \mapsto \Rcal \tensor \Rcal \textrm{   is injective} \nonumber \\
&&\gamma(\Delta(a\;\; mod (\Ical))=\Delta(a) mod(\Ical\tensor
\Ical)\nonumber
\end{eqnarray}
where $\Ical$ was a subideal of $\Rcal$ such that $\Ucal =
\Rcal/\Ical$. The last demand is equivalent to demanding that we
do not alter $\Delta$ at the level of $\Ucal$. Then we define
\begin{equation}
\Delta'(a) = \gamma \circ \Delta(a).
\end{equation}
We can be more explicit in the case of a condensate in a phase
described by a modified quantum double $F(H_m)\times\C H_{el}$, when
the right residual symmetry algebra $\Rcal$ is of the form
(\ref{eq:Rcalspec}) : $\Rcal = F(H_m/B)\tensor N$. We've seen that
$\Rcal$ is of this form for electric condensates, and defect
condensates. The action of $N$ on $B$ is well defined, because $N$
acts trivially on $B$. For every orbit $A_i$ in $G/B$ under the
action of $N$, pick a preferred element $g_i$. Now define a map
$\sigma: H_m/B\mapsto H_m$ such that $\sigma(g_i B)\in g_i B$. In
other words, $\sigma$ sends every left $B$ coset $g_iB$ into a
chosen element in $g_i B$. We demand that$^{17}$ for all $n\in N$
\begin{equation}
\sigma(n\cdot g_iB) =n\cdot\sigma(g_iB).
\end{equation}
Using $\sigma$, we define
\begin{eqnarray}
&& \gamma : (F(H_m)\times\C H_{el})\tensor (F(H_m/B)\tensor N)
\rightarrow \nonumber \\ && \quad \rightarrow (F(H_m)\times\C
H_{el}) \tensor
(F(H_m/B)\tensor N) \nonumber \\
&& \gamma (P_gn \tensor P_{g'B}n') = P_{\sigma(g'B)^{-1} g
\sigma(g'B)}n \tensor P_{g'B}n'.
\end{eqnarray}
It is straightforward but lengthy to verify that $\gamma
|_{\Delta(\Rcal)}$ is an algebra morphism$^{18}$. Thus we have a
coproduct
\begin{equation}
\Delta'(P_{gB}n)=\gamma\circ\Delta(P_{gB}n)=\sum_{g_i} P_{ \sigma
(g_iB)^{-1}gB} n \tensor  P_{g_iB} n.
\end{equation}

\subsection{An example: $\Acal=D(\overline{D_{2n}})$}

As an example, we consider a phase described by the quantum double
of the double cover of an even dihedral group:
$\Acal=D(\overline{D_{2n}})$. The group $\overline{D_{2n}}$ has
the following structure:
\begin{equation}
\overline{D_{2n}}=\{ r^k s^m : k=0,1,\ldots,2n-1 ; m=0,1 \}
\end{equation}
with $r^{2 n}= s^2 = -e$.

Now condense a single defect$^{19}$ $|s>$. The right residual and
unconfined symmetry algebras are
\begin{eqnarray}
&& \Rcal = F(\overline{D_{2n}}/(s))\tensor\C \overline{\Z_2\sd\Z_2} \\
&& \Ucal = D(\Z_2),
\end{eqnarray}
where $\overline{\Z_2\sd\Z_2} = \{ r^{kn}s^m : k=0,1,2,3 ; m=0,1
\}$.

The irreps of $\Rcal$ are given in Table 7, and the irreps of
$\overline{\Z_2\sd\Z_2}$, which occurs as a centralizer of two
orbits in $\overline{D_{2n}}/(s)$, are given in Table 8.

We write the basis of the irrep $\Omega^i_k$ as $|i>, |-i>$, where
$|i>$ corresponds to the defect $|r^i(s)>$ (remember that the
defects are defined modulo the condensate $<s>$). The action of
$\Rcal$ on this basis is set by
\begin{eqnarray}
&& P_{r^j(s)}\cdot |i> = \delta_{j,i} \nonumber \\
&& r^{tn} \cdot |i> = (-1)^{tk}|i> \quad t=0,1,2,3 \nonumber \\
&& s|i> = |-i>. \nonumber
\end{eqnarray}
We can write this compactly in one equation:
\begin{equation}
P_{r^j(s)}r^{tn}s^l |i>=\delta_{j,(1-2l)i}(-1)^{tk} |(1-2l)i>
\quad t\in\Z_4, l \in \Z_2.
\end{equation}

\begin{table}[t!b]
\begin{center}
\begin{tabular}{|c|c|c|c|}
\hline
                            &                           &               &               \\
Preferred           &   $e$                 &   $r^n$       & $r^i$             \\
element $g_A$       &                       &               & $i=0,1,\ldots,n-1$    \\
                            &                           &               &               \\
\hline
                            &                           &               &               \\
Orbit $A$                   &   $(s)=\{e,s,-e,-s\}$     &   $r^n(s)$    & $\{r^i(s),r^{-i}(s)\}$\\
in $\overline{D_{2n}}/(s)$  &                           &               &               \\
                            &                           &               &               \\
\hline
                            &                            &              &       \\
Normalizer $N_A$            &   $\overline{\Z_2\sd\Z_2}$=&$\overline{\Z_2\sd\Z_2}$=&    $\overline{\Z_2}$=  \\
                            &$\{r^{tn}s^k :         $   & $\{r^{tn}s^k:$ & $\{e,s,-e,-s\}$ \\
                            & $t\in\Z_4,k\in\Z_2\}$     & $t\in\Z_4,k\in\Z_2\}$ &       \\
                            &                           &               &               \\
\hline
                            &                           &               &               \\
Irrep of $N_A$          &   $\alpha_{k,l}$          &$\alpha_{k,l}$ & $\beta_m$         \\
                            &   $k\in\Z_4, l\in\Z_2$    &$k\in\Z_4, l\in\Z_2$ & $m\in\Z_4$   \\
                            &                           &               &               \\
\hline
                            &                           &               &               \\
Irrep of $\Omega$               &   $\Omega^0_{k,l}$        & $\Omega^n_{k,l}$ & $\Omega^i_m$   \\
                            &                           &               &               \\
\hline
                            &                           &               &               \\
Unconfined          &   $\Omega^0_{k,0}$        & $\Omega^n_{k,0}$ &                \\
irreps              &   $k=0,2$                 & $k=0,2$       &                   \\
                            &                           &               &               \\
\hline
\end{tabular} \vspace*{.2cm}
\end{center}
\label{tab:fixco} \caption{{\small  The irreps of
$\Rcal=F(\overline{D_{2n}}/(s)) \tensor \C \overline{\Z_2\sd
\Z_2}$.}}
\end{table}

\begin{table}
\begin{center}
\begin{tabular}{|c|c|c|c|c|c|c|c|c|}
\hline
                            &    &      &       &          &     &      &        &      \\
Element of                  &$e$ & $-e$ & $r^n$ & $r^{-n}$ & $s$ & $-s$ & $sr^n$ & $sr^{-n}$    \\
$\overline{\Z_2\sd\Z_2}$    &    &      &       &          &     &      &        &      \\
                            &    &      &       &          &     &      &        &      \\
\hline
                            &    &      &       &          &     &      &        &      \\
Irrep $\alpha_{k,l}$        & $1$ & $(-1)^k$ & $i^k$ & $i^{-k}$ &
$(-1)^li^k$ & $(-1)^li^{-k}$ & $(-1)^{l+k}$ &
$(-1)^m$ \\
$k\in\Z_4,\;l\in\Z_2$       & & & & & & & & \\
                            &    &      &       &          &     &      &        &      \\
\hline
\end{tabular} \vspace*{.2cm}
\end{center}
\label{tab:centsirrs} \caption{{\small  The irreps of
$\overline{\Z_2\sd\Z_2}$.}}
\end{table}

The left $(s)$ cosets in $\overline{D_{2n}}/(s)$ are
$r^i(s)=\{r^i,-r^i,r^is,-r^is : i=0,1,\ldots,2n-1\}$. To define
the coproduct, we must choose $\sigma(r^i(s))\in r^i(s)$ for every
$i$. If we choose $\sigma(r^i(s))=r^i$, then the coproduct
$\Delta'$ is coassociative. This can be traced back to the fact
that the $r^i$ form a group. Note that in general, it is not
possible to choose representatives of the cosets so that they form
a group.

Using $\Delta'$, we can determine the following fusion rule:
\begin{equation}
\Omega^i_k \tensor \Omega^j_l =
\Omega^{i+j}_{k+l}\oplus\Omega^{i-j}_{k+l}
\end{equation}

The unconfined irreps are given in Table 8. This fusion rules
implies, for example, that $\Omega^i_1$ and $\Omega^{n-i}_1$ can
fuse to $\Omega^n_2$, which is unconfined. Thus we have made a
hadronic composite.

If we choose $\sigma(r^i(s))=r^is$, the fusion rules are the same,
even though the coproduct is not coassociative. However, in this
case the definition of the particles is altered. Namely, if an
irrep $\Omega^i_k$ is present to the right, then $|j>$ in the
irrep $\Omega^j_l$ coming in from the left should be interpreted
as $|-j>$. This can be checked by applying the projection
operators:
\begin{equation}
P_{r^l(s)}\cdot(|j>\tensor|i>)=\delta_{l,-j+i} |j>\tensor|i>
\end{equation}
We have therefore discovered that our choice of coproduct
$\Delta'$ alters the meaning of the labels of our irreps, when
they are to the right of a confined excitation. The reason for
this is the following: we can determine the irreps of $\Rcal$, and
that gives us a finite set of labels. When we have one particle in
an irrep of $\Rcal$ in the system, its meaning is unambiguous. Now
if a confined excitation is present, then we know that the
residual symmetry algebra of the condensate $\phil$ to the left of
the excitation may be different from the right residual symmetry
algebra $\Rcal$. However, we still want to use the same labels for
particles to the left of the confined excitations, because they
are excitations of a symmetry algebra isomorphic to $\Rcal$. Thus
we are adding information to the labels of irreps of $\Rcal$,
namely we are defining their meaning when they appear to the left
of a confined excitation. They should not be considered as the
same particle: for example, $|i>|i>$ should not be interpreted as
the fusion of $|i>$ with itself. The correct interpretation is
that we have the confined excitation $|i>$ of $\Rcal$, and we
brought in a particle from the left, which under our choice of
coproduct is labelled by $|i>$.

In summary, for the $|s>$ condensate in $D(\overline{D_{2n}})$,
the fusion rules are independent of the choice of coproduct. We do
not expect this to be a general result (although in the cases
we've worked out the fusion rules don't depend on the choice of
coproduct). The \emph{physics} of the phase, on the other hand,
should not depend on our choice of coproduct, since this choice
boils down to a definition of our labels. An interesting follow up
on this research would be to study the influence of the choice of
coproduct on the fusion rules.

\newpage

\section*{Notes}

1 :  Modified quantum doubles are a slight variation on quantum
doubles. Their precise definition is discussed in Appendix
\ref{app:Hopf}.

2 :  We assume here that $G$ is simply connected. If it isn't, we
can take the universal covering group $\tilde{G}$ of $G$, and then
$H$ should be

replaced by the lift $\tilde{H}$ of $H$ in the universal covering
group $\tilde{G}$ in all formulae.

3 :  A comprehensive summary of the algebraic framework of Hopf
algebras, focusing on our specific needs is given in Appendix A. For
an extensive introduction to the use of Hopf algebras in the present
physical context we refer to \cite{dwpb1995}.

4 :  See Appendix A, and for an extensive background \cite{Majid}.

5 :  Actually, if we are braiding two indistinguishable electric
particles, then the braiding may give a phase factor. Under
half-braiding the wavefunction of the system picks up a phase factor
$e^{i2\pi s}$, where $s$ is the spin of the particles.

6 :  The difference with standard Riemann geometry is that one
doesn't assume that the metric is torsion free, i.e. the Christoffel
symbols $\Gamma_{\mu\nu}^{\kappa}$ are not required to satisfy
$\Gamma_{\mu\nu}^{\kappa}=\Gamma_{\nu\mu}^{\kappa}$. In the Cartan
formulation the spin connection and the vielbein are considered as
independent, while in Riemannian geometry they are linked by the no
torsion condition.

7 :  In a gauge theory, this amounts to picking a gauge such that
the Dirac string points `upwards' in the drawings. In the global
case, we orient the defect such that the frame dragging happens in
the upper half plane.

8 :  We thank  dr. Joost Slingerland for discussions on this point.
He studied this relation independently.

9 :  $\Rcal^{\perp}$ is the vector space consisting of all vectors
in $\Acal$ perpendicular to all vectors in $\Rcal$.

10 :  We can find such a basis of $\Acal$ using the Gram-Schmidt
orthogonalization procedure.

11 :  The kernel of a map is the set of elements which the irrep
maps to zero. These elements form an ideal $I$, meaning that if
$i\in I$ and $a\in\Rcal$, then $ia\in I$ and $ai\in I$.

12 :  To prove this, use $<kg_Ak^{-1}>=k<g_A>k^{-1}$ and
$N_{kg_Ak^{-1}}=kN_Ak^{-1}$.

13 :  Note that $g$ and $g^{-1}$ needn't be in the same conjugacy
class.

14 :  The dual of an algebra $\Acal$ is defined as
$\Acal^*=F(\Acal)$, the set of linear functions from $\Acal$ to
$\C$.

15 :   Different, strictly equivalent notations are frequently used
here: $ f_i = f_{h_i} = P_{h_i} $ . In particular we have the
identities $ f_i (x) = \delta_{h_i} (x) = \delta_{{h_i} x^{-1}} (e)
= \delta_{{h_i}, x} $ , the last delta being the usual Kronecker
delta.

16 :  We came up with this structure to deal with phases with
inversion or reflection symmetries. It turns out to be a special
case of a bicrossproduct of Hopf algebras, see e.g. \cite{Majid}.

17 :  It is not at all clear that this can always be achieved, even
though it is possible in the simple cases we've looked at.

18 :  $\gamma$ is not an algebra morphism of $D(H)\tensor\Rcal$, it
is only an algebra morphism when restricted to $\Delta(\Rcal)$.

19 :  We take even dihedral groups, because in
$D(\overline{D_{2n+1}})$, condensing $|s>$ yields $\Ucal=\C e$,
which is slightly less interesting, but note that we can still study
hadrons! Only the hadron must be in the trivial irrep, since that is
the only unconfined irrep.

\bibliographystyle{unsrt}

\newpage

\begin{thebibliography}{35}
\bibitem{Bais:2002pb} F.A. Bais, B.J. Schroers, J.K. Slingerland, Broken quantum
symmetry and confinement phases in planar physics, Phys. Rev. Lett.
89 (2002) 181601

\bibitem{Bais:2002ny} F.A. Bais, B.J. Schroers, J.K. Slingerland, Hopf symmetry
breaking and confinement in (2+1)-dimensional gauge theory, JHEP 05
(2003) 068

\bibitem{mbnematic:2006} C.J.M. Mathy, F.A. Bais, Nematic phases and the breaking of
double symmetries, arXiv:cond-mat/0602109, 2006.

\bibitem{bmmelting:2006} F.A. Bais, C.J.M. Mathy, Defect mediated melting and the
breaking of quantum double symmetries, arXiv:cond-mat/0602101, 2006.

\bibitem{Bais:1991pe} F.A. Bais, P. van Driel, M. de Wild Propitius, Quantum
symmetries in discrete gauge theories, Phys. Lett. B280 (1992) 63-70

\bibitem{dwpb1995} M. de Wild Propitius, F.A. Bais, Discrete gauge theories, in: G.
Semenoff and L. Vinet (Ed.), Particles and Fields, CRM Series in
Mathematical Physics, Springer Verlag, New York, 1998, p. 353-439.

\bibitem{Moore:1991ks} G.W. Moore, N. Read, NonAbelians in the Fractional Quantum Hall
Effect, Nucl. Phys. B360 (1991) 362-396

\bibitem{Read:1998ed} N. Read, E. Rezayi, Beyond paired Quantum Hall states:
parafermions and incompressible states in the first excited Landau
Level, Phys. Rev. B59 (1999) 8084.

\bibitem{Nayak:1995kx} C. Nayak, F. Wilczek, Quantum Hall states of
high symmetry, Nucl. Phys. B450 (1995) 558-568.

\bibitem{Slingerland:2001ea} J.K. Slingerland, F.A. Bais, Quantum groups and nonabelian
braiding in Quantum Hall systems, Nucl. Phys. B612 (2001) 229-290.

\bibitem{Fendley:2005yy} P. Fendley, E. Fradkin, Realizing non-Abelian statistics, Phys.
Rev. B72 (2005) 024412.

\bibitem{Bais:1998yn} F.A. Bais, N.M. Muller, Topological field theory and the quantum
double of SU(2), Nucl. Phys. B530 (1998) 349-400.

\bibitem{Kitaev:1997wr} A.Y. Kitaev, Fault-tolerant quantum computation by anyons, Ann.
Phys. 303 (2003) 2-30.

\bibitem{Preskill:1997ds} J. Preskill, Reliable Quantum Computers, Proc. Roy. Soc. Lond.
A454 (1998) 385-410.

\bibitem{Freedman:2000rc} M.H. Freedman, A. Kitaev, Z. Wang, Simulation of topological
field theories by quantum computers, Commun. Math. Phys. 227 (2002)
587-603.

\bibitem{Dennis:2001nw} E. Dennis, A. Kitaev, A. Landahl, J. Preskill, Topological
quantum memory, J. Math. Phys. 43 (2002) 4452-4505.

\bibitem{Bais:1980fm} F.A. Bais, Flux metamorphosis, Nucl. Phys. B170 (1980) 32-43.

\bibitem{Majid} S. Majid, Foundations of Quantum Group Theory, Cambridge
University Press, 1995.

\bibitem{Propitius} M. de Wild Propitius, Topological interactions in broken gauge
theories, PhD thesis, Universiteit van Amsterdam, 1995.

\bibitem{Kleinert2} H. Kleinert, Gauge Fields in Condensed Matter, Vol. I: Superflow
and Vortex Lines, Disorder Fields, Phase Transitions, World
Scientific, Singapore, 1989.

\bibitem{Katanaev} M.O. Katanaev, Geometric theory of defects, PHYS-USP 48(7)
(2005) 675-701.

\bibitem{Khazan} M.V. Khazan, Analog of the Aharonov-Bohm effect in superfluid
He3-A, JETP Lett. 41 (1985) 486.

\bibitem{Furtado:2000} C. Furtado, V.B. Bezerra, F. Moraes, Berry's quantum phase in
media with dislocations, Europhys. Lett. 52 (2000), 1.

\bibitem{McGraw} P. McGraw, Global analogue of Cheshire charge, Phys. Rev. D50
(1994) 952-961.

\bibitem{Verlinde} E. Verlinde, A Note on Braid Statistics and the Non-Abelian
Aharonov-Bohm Effect, in: S. Dal (Ed.), International Colloquium on
Modern Quantum Field Theory, Bombay 1990, World SCientific,
Singapore, 1991.

\bibitem{March-Russell:1991az} J. March-Russell, J. Preskill, F. Wilczek, Internal frame
dragging and a global analog of the Aharonov-Bohm effect, Phys. Rev.
Lett. 68 (1992) 2567-2571.

\bibitem{Bais:1980nn} F.A. Bais, The topology of monopoles crossing a phase boundary,
Phys. Lett. B98 (1981) 437.

\bibitem{Jin} M. Greiner, C.A. Regal, D.S. Jin, Fermionic condensates, in: AIP
Conf. Proc., vol. 77-, 2005, p. 209-217.

\bibitem{Lubensky:bent-core} T.C. Lubensky, L. Radzihovsky, Theory of bent-core
liquid-crystal phases and phase transitions, Phys. Rev. E 66 (2002)
031704.

\bibitem{Lubensky:rg} L. Radzihovsky, T.C. Lubensky, Fluctuation-driven 1st-order
isotropic-to-tetrahedratic phase transition, Europhys. Lett. 54(2)
(2001) 206-212.

\bibitem{Salt} J.A. Salthouse, M.J. Ware, Point group character tables and
related data, Cambridge University Press, 1972.

\bibitem{Niori} T. Niori, T. Sekine, J. Watanabe, T. Furukawa, H. Takezoe, J.
Mater. Chem. 6 (1996) 1231.

\bibitem{Kassel} C. Kassel, Quantum groups, Springer Verlag, Berlin, 1995.

\bibitem{Koornwinder:loccpt} T.H. Koornwinder, N.M. Muller, The quantum double of a (locally)
compact group, J. of Lie Theory 7 (1997) 101-120.

\bibitem{Koornwinder:cpt} T.H. Koornwinder, F.A. Bais, N.M. Muller, Tensor product
representations of the quantum double of a compact group, Comm. in
Math. Phys. 198 (1998) 157-186.

\end{thebibliography}

\end{document}